\documentclass[letterpaper,aps,showkeys,showpacs,amsmath,amssymb,pre,groupedaddress,12pt,preprint,tightenlines]{revtex4}

\usepackage{amsfonts}
\usepackage{graphicx}
\usepackage{natbib}
\usepackage{dcolumn}
\usepackage{bm}
\usepackage[letterpaper,left=0.8in,right=0.8in,top=0.7in,bottom=0.8in]{geometry}
\usepackage{rotating}
\usepackage{longtable}

\newcommand{\beq}{\begin{equation}}
\newcommand{\eeq}{\end{equation}}
\newcommand{\bea}{\begin{eqnarray}}
\newcommand{\eea}{\end{eqnarray}}
\newcommand{\ba}{\begin{array}}
\newcommand{\ea}{\end{array}}
\newcommand{\bi}{\begin{itemize}}
\newcommand{\ei}{\end{itemize}}
\newcommand{\ben}{\begin{enumerate}}
\newcommand{\een}{\end{enumerate}}

\long\def\symbolfootnote[#1]#2{\begingroup\def\thefootnote{\fnsymbol{footnote}}\footnote[#1]{#2}\endgroup}

\begin{document}

\preprint{}

\title{Multinetwork of International Trade:\\A Commodity-Specific Analysis}

\author{Matteo Barigozzi}
\affiliation{ECARES - Universit\'{e} Libre de Bruxelles, 50 Avenue
F.D. Roosevelt CP 114, 1050 Brussels, Belgium.\\ Tel: +32 (0)2 650 33
75 . E-mail: matteo.barigozzi@ulb.ac.be}

\author{Giorgio Fagiolo}
\affiliation{Sant'Anna School of Advanced Studies,\\ Laboratory of Economics and
Management, Piazza Martiri della Libert\`{a} 33, I-56127 Pisa, Italy.\\ Tel:
+39-050-883359 Fax: +39-050-883343. E-mail: giorgio.fagiolo@sssup.it}

\author{Diego Garlaschelli}
\affiliation{Sant'Anna School of Advanced Studies,\\ Laboratory of Economics and
Management, Piazza Martiri della Libert\`{a} 33, I-56127 Pisa, Italy.\\
E-mail: diego.garlaschelli@sssup.it}


\begin{abstract}
\noindent 
We study the topological properties of the multinetwork of commodity-specific trade relations among world
countries over the 1992-2003 period, comparing them with those of the aggregate-trade network, known in the
literature as the international-trade network (ITN). We show that link-weight distributions of commodity-specific
networks are extremely heterogeneous and (quasi) log normality of aggregate link-weight distribution
is generated as a sheer outcome of aggregation. Commodity-specific networks also display average connectivity,
clustering, and centrality levels very different from their aggregate counterpart. We also find that ITN
complete connectivity is mainly achieved through the presence of many weak links that keep commodity-specific
networks together and that the correlation structure existing between topological statistics within each
single network is fairly robust and mimics that of the aggregate network. Finally, we employ cross-commodity
correlations between link weights to build hierarchies of commodities. Our results suggest that on the top of a
relatively time-invariant ``intrinsic" taxonomy (based on inherent between-commodity similarities), the roles
played by different commodities in the ITN have become more and more dissimilar, possibly as the result of
an increased trade specialization. Our approach is general and can be used to characterize any multinetwork
emerging as a nontrivial aggregation of several interdependent layers.
\end{abstract}

\keywords{Weighted directed networks; International trade network;
Multinetworks; Commodity-specific trade; Econophysics.}

\pacs{89.75.-k, 89.65.Gh, 87.23.Ge, 05.70.Ln, 05.40.-a}

\maketitle

\section{Introduction} \label{Sec:Intro}
The past decade has seen an increasing interest in the study of
international-trade issues from a complex-network perspective
\citep{LiC03,SeBo03,Garla2004,Garla2005,serrc07,Bhatta2007a,Bhatta2007b,
Garla2007,Tzekina2008,Fagiolo2008physa,Fagiolo2008acs,Fagiolo2009pre}. Existing
contributions have attempted to investigate the time-evolution of the
topological properties of the aggregate International Trade Network (ITN), aka
the World Trade Web (WTW), defined as the graph of all import/export
relationships between world countries in a given year.

Two main approaches have been employed to address this issue. In the first one,
the ITN is viewed as a binary graph where a (possibly directed) link is either
present or not according to whether the value of the associated trade flow is
larger than a given threshold \citep{SeBo03,Garla2004,Garla2005}. In the second
one, a \textit{weighted-network approach} \citep{Barrat2004pnas,Barthelemy2005}
to the study of the ITN has been used, i.e. links between countries are
weighted by the (deflated) value of imports or exports occurred between these
countries in a given time interval
\citep{LiC03,Bhatta2007a,Bhatta2007b,Garla2007,Fagiolo2008physa,Fagiolo2009pre}.
In most cases, a symmetrized version of the ITN has been studied, where only
undirected trade flows are considered and one neglects ---in a first
approximation--- the importance of directionality of trade flows.

Such studies have been highlighting a wealth of fresh stylized facts
concerning the architecture of the ITN, how they change through
time, how topological properties correlate with country
characteristics, and how they are predictive of the likelihood that
economic shocks might be transmitted between countries
\citep{KaliReyes2007}. However, they all consider the web of world
trade among countries at the aggregate level, i.e. links represent
total trade irrespective of the commodity actually traded
\footnote{See Refs. \cite{ReichardtWhite2007} and
\cite{Lloyd_etal2009} for exceptions. Refs.
\citep{Hidalgo_etal_2007_Science,Hidalgo_Hausmann_2009} also employ
commodity-specific data to build a network view of economic
development where one analyzes a tripartite graph, linking countries
to the products they export and the capabilities needed to produce
them. Unlike the present study, however, they do not explicitly
consider the web of trade relations between any pair of countries.}.
Here we take a commodity-specific approach and we unfold the
aggregate ITN in many layers, each one representing import and
export relationships between countries for a given commodity class
(defined according to standard classification schemes, see below).

More precisely, we employ data on bilateral trade flows taken from the United
Nations Commodity Trade Database to build a multi-network of international
trade. A multi-network \citep{Wasserman1994} is a graph where a finite,
constant set of nodes (world countries) are connected by edges of different
colors (commodities). Any two countries might then be connected by more than
one edge, each edge representing here a commodity-specific flow of
imports/exports. As our data span a 12-year interval, $N=162$ countries and
$C=97$ commodities, we therefore have a sequence of 12 international-trade
multi-networks (ITMNs), where between any pair of the $N$ countries there may
be at most $C$ edges. Each ITMN can then be viewed in its entirety or also as the
juxtaposition of $C=97$ commodity-specific networks, each modeled as a weighted directed network.
We weight a link from country $i$ to $j$ by the (properly rescaled) value of $i$'s
exports to $j$, and, in general, the link from $i$ to $j$ is different from the link from $j$ to $i$.

The multi-network setup allows us to ask novel questions related to the
structural properties of the ITN. For example: To what extent do topological
properties of the aggregate ITN depend on those of the commodity-specific
networks? Are trade architectures heterogeneous across commodity-specific
networks? How do different topological properties correlate within each
commodity-specific network, and how do the same topological property
cross-correlates across commodity-specific network? How do countries perform in
different commodity-specific networks as far as their topological properties
are concerned (i.e. centrality, clustering, etc.)? Is it possible to build
correlation-based distances among commodities and build taxonomies that account
for ``intrinsic'' factors (inherent similarity between commodities as described
in existing classification schemes) as well as for ``revealed'' factors
(determined by the actual pattern of trades)?

In this paper we begin answering these questions. Our preliminary results show
that commodity-specific networks are extremely heterogeneous as far as
link-weight distributions are concerned and that the (quasi) log-normality of
aggregate link-weight distribution is generated as a sheer outcome of
aggregation of statistically dissimilar commodity-specific distributions.
Commodity-specific networks also display average connectivity, clustering and
centrality levels very different from their aggregate counterparts. We also
study the connectivity patterns of commodity-specific networks and find that
complete connectivity reached in the aggregate ITN is mainly achieved through
the presence of many weak links that keep commodity-specific networks together,
whereas strong trade links account for tightly interconnected clubs of
countries that trade with each other in all commodity networks. We also show
that, despite a strong distributional heterogeneity among commodity-specific
link-weight distributions, the correlation structure existing between
topological statistics within each single network is fairly robust and mimics
that of the aggregate network. Furthermore, we find that cross-commodity
correlations of the same statistical property are almost always positive,
meaning that on average large values of node clustering and centrality in a
commodity network imply large values of that statistic also in all other
commodity networks. Finally, we introduce a general method to characterise hierarchical dependencies among layers in multi-networks, and we use it to
compute cross-commodity correlations. We exploit these correlations between
link weights to explore the possibility of building taxonomies of commodities.
Our results suggest that on the top of a relatively time-invariant
``intrinsic'' taxonomy (based on inherent between-commodity similarities), the
roles played by different commodities in the ITN have become more and more
dissimilar, possibly as the result of an increased trade specialization.

The rest of the paper is organized as follows. Section \ref{Sec:Data_and_Defs}
describes the database, explains the methodology employed to build the ITMNs
and defines the basic topological statistics employed in the analysis. Sections
\ref{Sec:Results} and \ref{Sec:sec_corr} report our main results. Concluding
remarks are in Section \ref{Sec:Conclusions}.

\section{Data and Definitions} \label{Sec:Data_and_Defs}

\subsection{Data} \label{SubSec:Data}

We employ data on bilateral trade flows taken from the United Nations Commodity
Trade Database (UN-COMTRADE; see \texttt{http://comtrade.un.org/}). We build a
balanced panel of $N=162$ countries for which we have commodity-specific
imports and exports flows from 1992 to 2003 ($T=12$ years) in current U.S.
dollars. Trade flows are reported for $C=97$ (2-digit) different commodities,
classified according to the Harmonized System 1996 (HS1996; see Table
\ref{Tab:HS} and \texttt{http://www.wcoomd.org/})\footnote{Since, as always
happens in trade data, exports from country $i$ to country $j$ are reported
twice (according to the reporting country --- importer or exporter) and
sometimes the two figures do not match, we follow Ref. \citep{Feenstra2005data}
and only employ import flows. For the sake of exposition, however, we follow
the flow of goods and we treat imports from $j$ to $i$ as exports from $i$ to
$j$.}.

\subsection{The International-Trade Multi-Network} \label{SubSec:ITMN}

We employ the database to build a time sequence of weighted, directed
multi-networks of trade where the $N$ nodes are world countries and directed
links represent the value of exports of a given commodity in each year or wave
$t=1992,\dots,2003$. As a result, we have a time sequence of $T$ multi-networks
of international trade, each characterized by $C$ layers (or links of $C$
different colors). Each layer $c=1,\dots,C$ represents exports between
countries for commodity $c$ and can be characterized by a $N\times N$ weight
matrix $X_t^c$. Its generic entry $x_{ij,t}^c$ corresponds to the value of
exports of commodity $c$ from country $i$ to country $j$ in year $t$. We
consider directed networks, therefore in general $x_{ij,t}^c\neq x_{ji,t}^c$.
The aggregate weighted, directed ITN is obtained by simply summing up all
commodity-specific layers. The entries of its weight matrices $X_t$ will read:

\beq
x_{ij,t}= \sum_{c=1}^{C}{x_{ij,t}^c}.
\eeq

In order to compare networks of different commodities at a given time $t$, and
to wash-away trend effects, we re-scale all commodity-specific trade flows by
the total value of trade for that commodity in each given year. This means that
in what follows we shall study the properties of the sequence of
international-trade multi-networks (ITMNs) where the generic entry of the
weight matrix is defined as:

\beq \label{Eq:WeightsSpecific}
w_{ij,t}^c=\frac{x_{ij,t}^c}{\sum_{h=1}^N\sum_{k=1}^N x_{hk,t}^c}. \eeq
Therefore, the directed $c$-commodity link from country $i$ to country $j$ in
year $t$ is weighted by the ratio between exports from $i$ to $j$ of $c$ to
total year-$t$ trade of commodity $c$.

Accordingly, the generic entry of the aggregate-ITN weight matrix is re-scaled
as:

\beq \label{Eq:DivByExp} w_{ij,t}=\frac{x_{ij,t}}{\sum_{h=1}^N\sum_{k=1}^N
x_{hk,t}}. \eeq

Commodity-specific adjacency (binary) matrices $A_{t}^c$ are obtained from
weighted ones by simply setting $a_{ij,t}^c=1$ if and only if the corresponding
weight is larger than a given time- and commodity-specific threshold
$\underline{w}_t^c$. Unless explicitly noticed, we shall set
$\underline{w}_t^c=0$.

Before presenting a preliminary descriptive analysis of the data, two issues
are in order. First, most of our analysis below will focus on year 2003 for the
sake of simplicity. We employ a panel description in order to keep a fixed-size
country network and avoid difficulties related to across-year comparison of
topological measures, when required. Of course, accounting for entry/exit of
countries in the network may allow one to explore hot issues in international
trade literature as the relative importance of intensive and extensive margins
of trade from a commodity specific approach
\citep{HummelsKlenow2005,DeBene2008}. Although all our results seem to be
reasonably robust in alternative years, a more thorough comparative-dynamic
analysis is the next point in our agenda. Second, in order to correctly account
for trend effects, one should deflate commodity-specific trade flows by its
industry-specific deflator, which unfortunately is not available for all
countries. That is why we have chosen to remove trend effects and scale trade
flows by total commodity-specific trade in that year.

\subsection{Commodity Space} \label{SubSec:CommoditySpace}

\begin{figure}[t]
\begin{center}
\begin{minipage}{7cm}
\includegraphics[width=7cm,height=7cm]{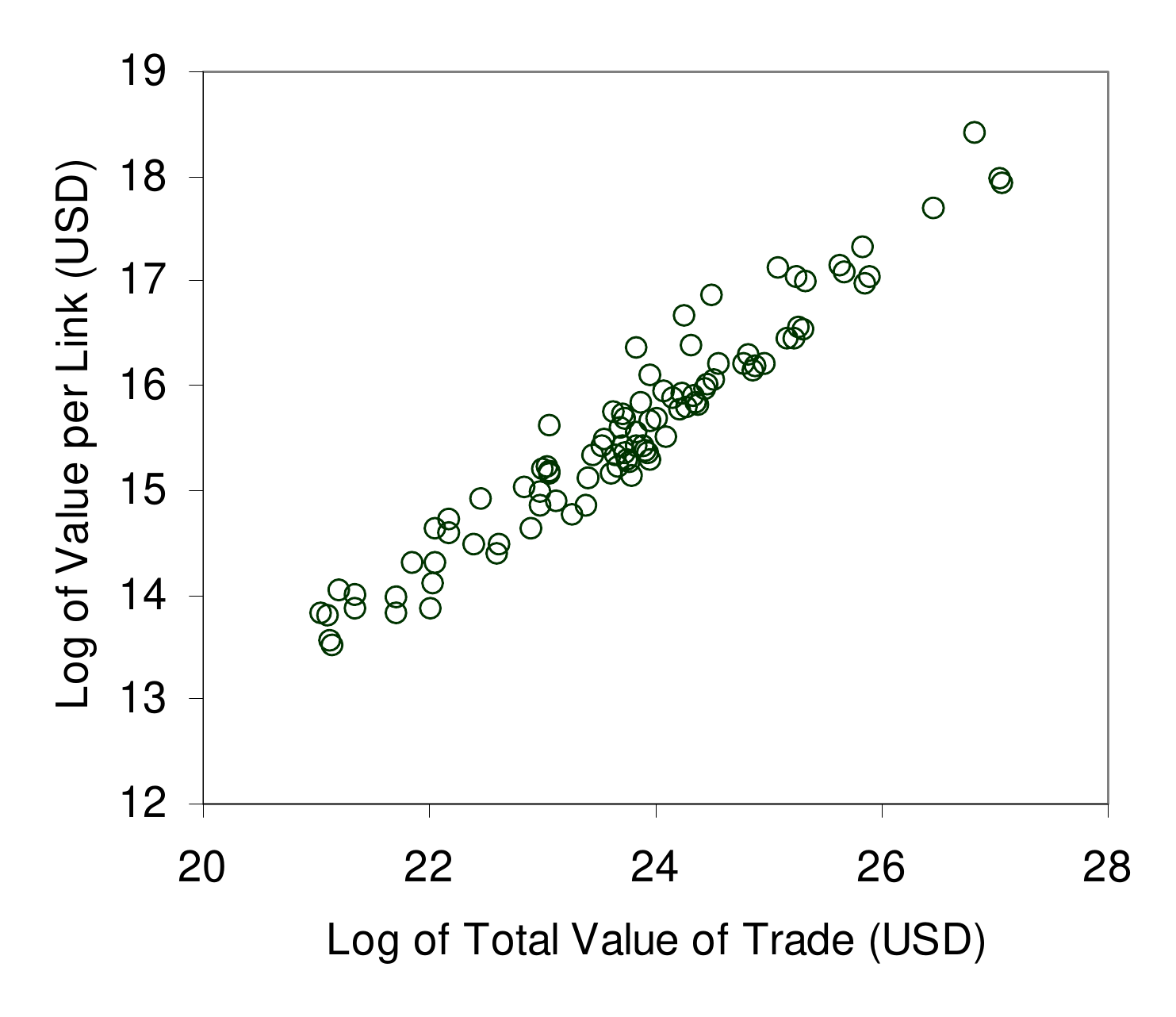}
\caption{{\label{fig:value_vs_valueperlink}}Scatter plot of total-trade value
vs. trade value per link of all 96 commodity classes in year 2003. Natural
logarithms on both axes.}
\end{minipage}
\begin{minipage}{0.5cm}
\
\end{minipage}
\begin{minipage}{7cm}
\includegraphics[width=7cm,height=7cm]{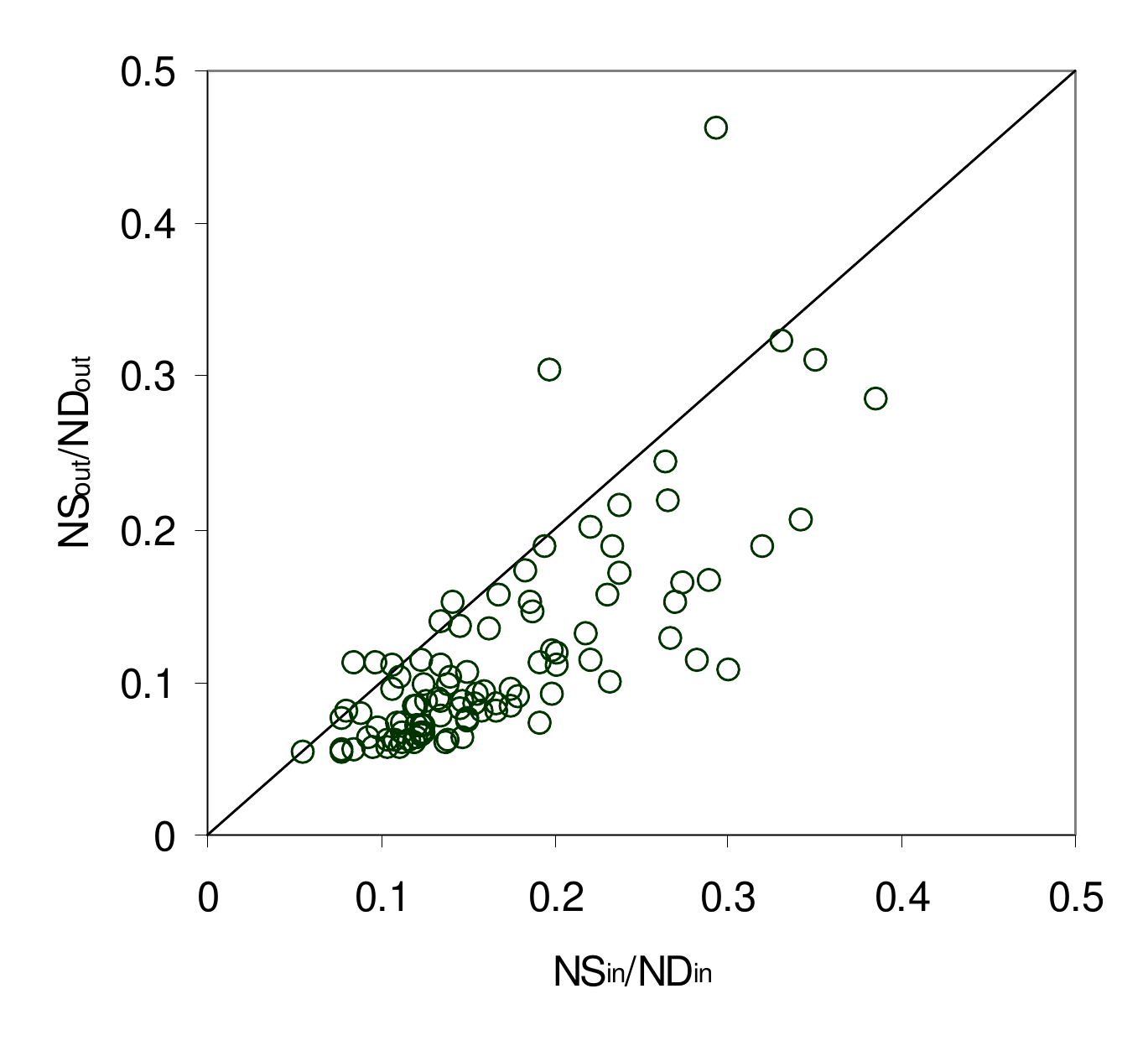}
\caption{{\label{fig:nsndin_nsndout}}Node in-strength per inward link vs node out-strength
per outward link of all 96 commodity classes in year 2003.}
\end{minipage}
\end{center}
\end{figure}

One of the aims of the paper, as mentioned, is to assess the across-commodity
heterogeneity of commodity-specific networks in terms of their topological
properties, as compared to those of the aggregate network. For the sake of
exposition, we shall focus, when necessary, on the most important commodity
networks. Table \ref{Tab:Top14} shows the ten most-traded commodities in 2003,
ranked according to the total value of trade. Notice that they account,
together, for 56\% of total world trade and that the 10 most-traded commodities
feature also the highest values of trade-value per link (i.e. ratio between
total trade and total number of links in the commodity-specific network).
Indeed, total trade value and trade-value per link of commodities are
positively correlated (see Figure \ref{fig:value_vs_valueperlink}), as are
total-trade value and network density (with a correlation coefficient of 0.52).
In addition to those trade-relevant 10 commodities, we shall also focus on
other 4 classes (cereals, cotton, coffee/tea and arms), which are less traded
but more relevant in economics terms. The 14 commodities considered account
together for 57\% of world trade in 2003.

\subsection{Topological Properties} \label{SubSec:TopoProp}

In the analysis below we shall focus on the following topological measures to
characterize trade networks and to compare them across commodities:

\begin{itemize}

\item Density (d): Network density is defined as the share of existing to
    maximum possible links in the binary $N\times N$ matrix.

\item Node in-degree ($ND_{in}$) and out-degree ($ND_{out}$): measure the
    number of countries from (respectively, to) which a given node imports
    (respectively, exports).

\item Node in-strength ($NS_{in}$) and out-strength ($NS_{out}$): Account
    for the share of country's total imports (respectively, exports) to
    world total commodity trade; more generally, in-strength (respectively,
    out-strength) is defined as the sum of all weights associated to inward
    (respectively, outward) links of a node. Node strength (NS) is simply
    defined as the sum of $NS_{in}$ and $NS_{out}$. Interesting statistics
    are also the ratios $NS_{in}/ND_{in}$ (average share of import per
    import partner) and $NS_{out}/ND_{out}$ (average share of export per
    export partner).

\item Node average nearest-neighbor strength (ANNS): measures the average
    NS of all the partners of a node. ANNS can be declined in four
    different ways, according to which one only considers the average
    $NS_{in}$ or $NS_{out}$ of import or export partners. Hence,
    $ANNS_{in-out}$ (respectively, $ANNS_{in-in}$) account for the average
    values of exports (respectively, imports) of countries from which a
    given node imports; similarly, $ANNS_{out-in}$ (respectively,
    $ANNS_{out-out}$) represent the average values of imports
    (respectively, exports) of countries to which a given node exports;

\item Node weighted clustering coefficient ($WCC_{all}$): proxies the
    intensity of trade triangles with that node as a vertex, where each
    edge of the triangle is weighted by its link weight
    \citep{Saramaki2006}. In weighted directed networks, one might
    differentiate across four types of directed triangles and compute four
    different types of clustering coefficients \citep{Fagiolo2007pre}: (i)
    $WCC_{mid}$, measuring the intensity of trade triangles where node $i$
    (the middleman) imports from $j$ and exports to $h$, which in turn
    imports from $j$; (ii) $WCC_{cyc}$, measuring the intensity of trade
    triangles where nodes $i$, $j$ and $h$ create a cycle; (iii)
    $WCC_{in}$, accounting for triangles where node $i$ imports from both
    $j$ and $h$; and (iv) $WCC_{out}$, accounting for triangles where node
    $i$ exports to both $j$ and $h$.

\item Node weighted centrality (WCENTR): measures the importance of a node
    in a network. Among the many suggested measures of node centrality
    \citep{CentralityIndices}, we employ here a version of Bonacich
    eigenvector-centrality suited to weighted-directed networks
    \citep{Bonacich2001}. It assigns relative scores to all nodes in the
    network based on the principle that connections to high-scoring nodes
    contribute more to the score of the node in question than equal
    connections to low-scoring nodes.

\end{itemize}

In addition to the above topological statistics, we also study the
distributions of link weights (both across commodity networks and in the
aggregate). Finally, we shall explore patterns of binary connectivity by
studying the properties (e.g. size and composition ) of the largest connected
component \footnote{A connected component of an undirected graph is a subgraph
in which any two vertices are connected to each other by paths, and to which no
more vertices or edges can be added while preserving its connectivity. That is,
it is a maximal connected subgraph. In directed graphs, one must firstly define
what it means for two nodes to be connected. We shall employ two different ways
to define wether any two nodes in the binary directed graph are connected.
According to the weaker one, any two nodes are connected if there is at least
one directed link between the two. The stronger one assumes two nodes to be
connected if there is a bilateral link between them.}.

\section{Topological Properties of Commodity-Specific Networks} \label{Sec:Results}

\subsection{Commodity-specific sample moments of topological properties} \label{SubSec:Moments}
We begin with a comparison of sample moments (mean and standard deviation) of
the relevant link and node statistics across different commodities. We compare
sample moments to those of the aggregate network to assess the degree of
heterogeneity of commodity networks and single out those that behave
excessively differently from the aggregate counterpart.

Table \ref{Tab:Averages} reports the density of the 14 most relevant
commodities, together with the mean and standard deviation of a few
link-weight and node-statistic distributions as described in Section
\ref{SubSec:TopoProp}. Notice that, as compared to the aggregate
network, all commodity-specific networks display larger average link
weights, shares of export/link and import/link, as well as overall
clustering. This means that connectivity and clustering patterns of
the commodity-specific trade networks are more intense than their
aggregate counterpart once one washes away the relative composition
of world trade. Conversely, by definition, all commodity-specific
densities are smaller than in the aggregate. Among the 14 most
relevant commodities, however, there appears to be a marked
heterogeneity. For example, arms (code 93) display a relatively low
density but a very strong average link weight and the largest import
and export per link shares and clustering. Cereals, on the other
hand, display a relatively small density as compared to the
aggregate, but exhibit a very large average link weight and shares
of import per inward link. The latter is larger than the average
shares of export per outward link, a result that generalizes for
almost all commodity-specific networks, see Figure
\ref{fig:nsndin_nsndout}. Larger shares of exports per outward link
are associated to larger shares of imports per inward link, but the
relative weight of imports dominates. This means that on average
countries tend to have, irrespective of the commodity traded and its
share on world market, more intensive import relations than export
ones (see also subsection \ref{SubSec:CorrAcross}).

\begin{figure}[t]
\begin{center}
\includegraphics[width=7cm,height=7cm]{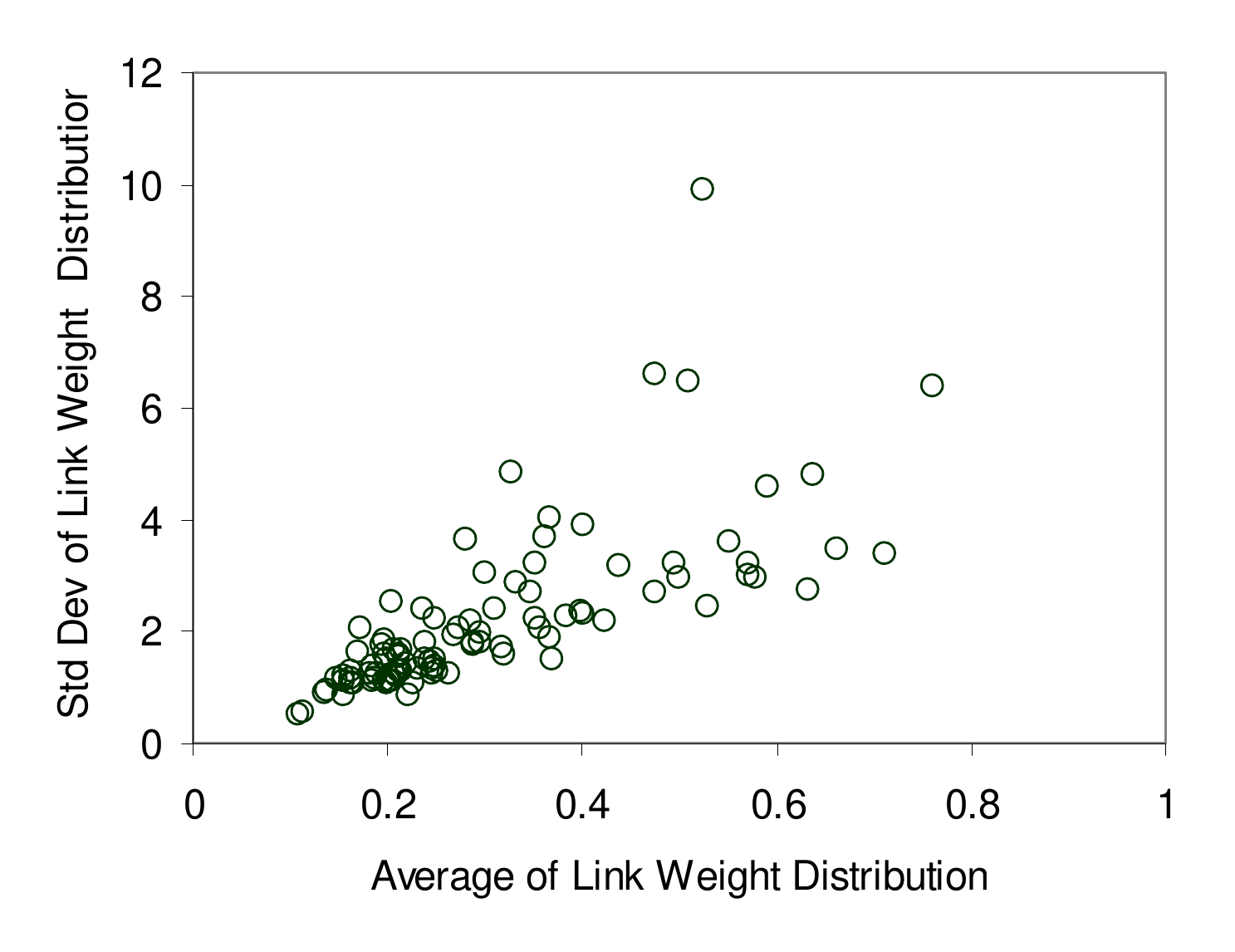}
\caption{{\label{fig:ave_stdev_linkweights}}Average vs. standard deviation of link-weight
distribution in 2003.}
\end{center}
\end{figure}

Another fairly general evidence regards the scaling between average and
standard deviation in link and node distributions. There appears to be a
positive relation between average and standard deviation of node and link
statistics (see Figure \ref{fig:ave_stdev_linkweights} for the example of link
weights), suggesting that within each commodity-specific network larger trade
intensities and clustering levels are gained at the expense of a much strong
heterogeneity in the country-distributions of such topological features.

To conclude this preliminary analysis, we report some results on the directed
clustering patterns observed across commodity networks. Following
\citep{Fagiolo2007pre}, we compute the percentage of directed trade triangles
of different types that each country forms with their partners (see Table
\ref{Tab:ClustPerc}). Note that in the aggregate network there is a slight
preponderance of out-type triangles (patterns where a country exports to two
countries that are themselves trade partners). Conversely, commodity-specific
networks are characterized also by a large fraction of in-type clustering
patterns (a country importing from two countries that are themselves trade
partners),except coffee and precious metals for which out-type clustering is
more frequent. The other two types of clustering patterns (cycle and middlemen)
are much less frequent.

\subsection{Distributional Features of Topological Properties} \label{SubSec:Distributions}
The foregoing results on average-dispersion scaling and heterogeneity across
commodity networks suggests that the overall evidence on aggregate trade
topology may be the result of extremely heterogeneous networks. For example,
previous studies on other data \citep{Fagiolo2009pre,GledData2002} have
highlighted the pervasiveness of log-normal shapes as satisfactory proxies to
describe the link- and node-distributions of aggregate link-weights, strength,
clustering and so on, in symmetrized versions of the ITN. Only node centrality
measures (computed using the notion of random-walk betweenness centrality, see
Ref. \citep{FisherVega2006}) seemed to display power-law shaped behavior.

\begin{figure}[t]
\begin{center}
\begin{minipage}{7cm}
\includegraphics[width=7cm]{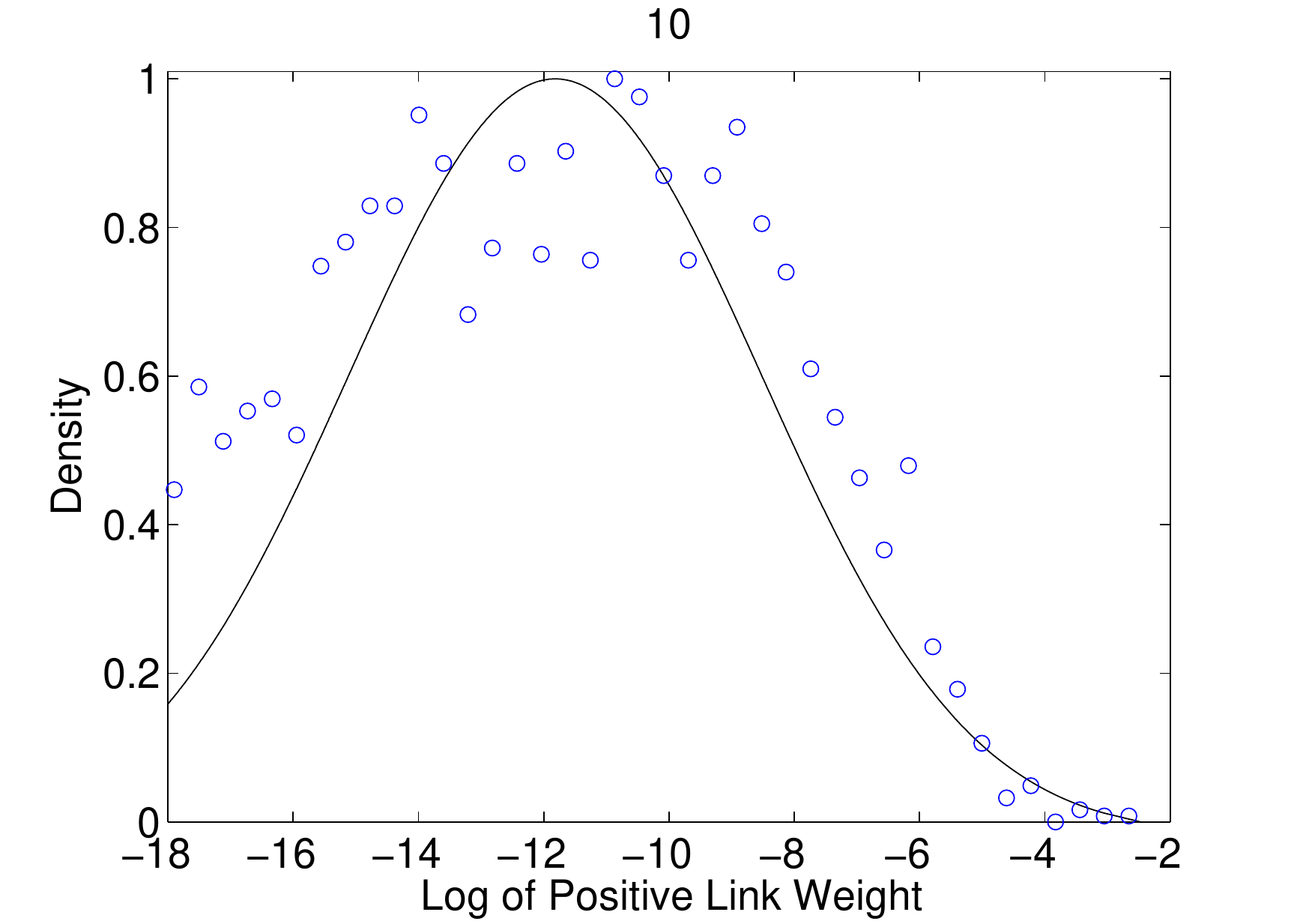}
\end{minipage}
\begin{minipage}{7cm}
\includegraphics[width=7cm]{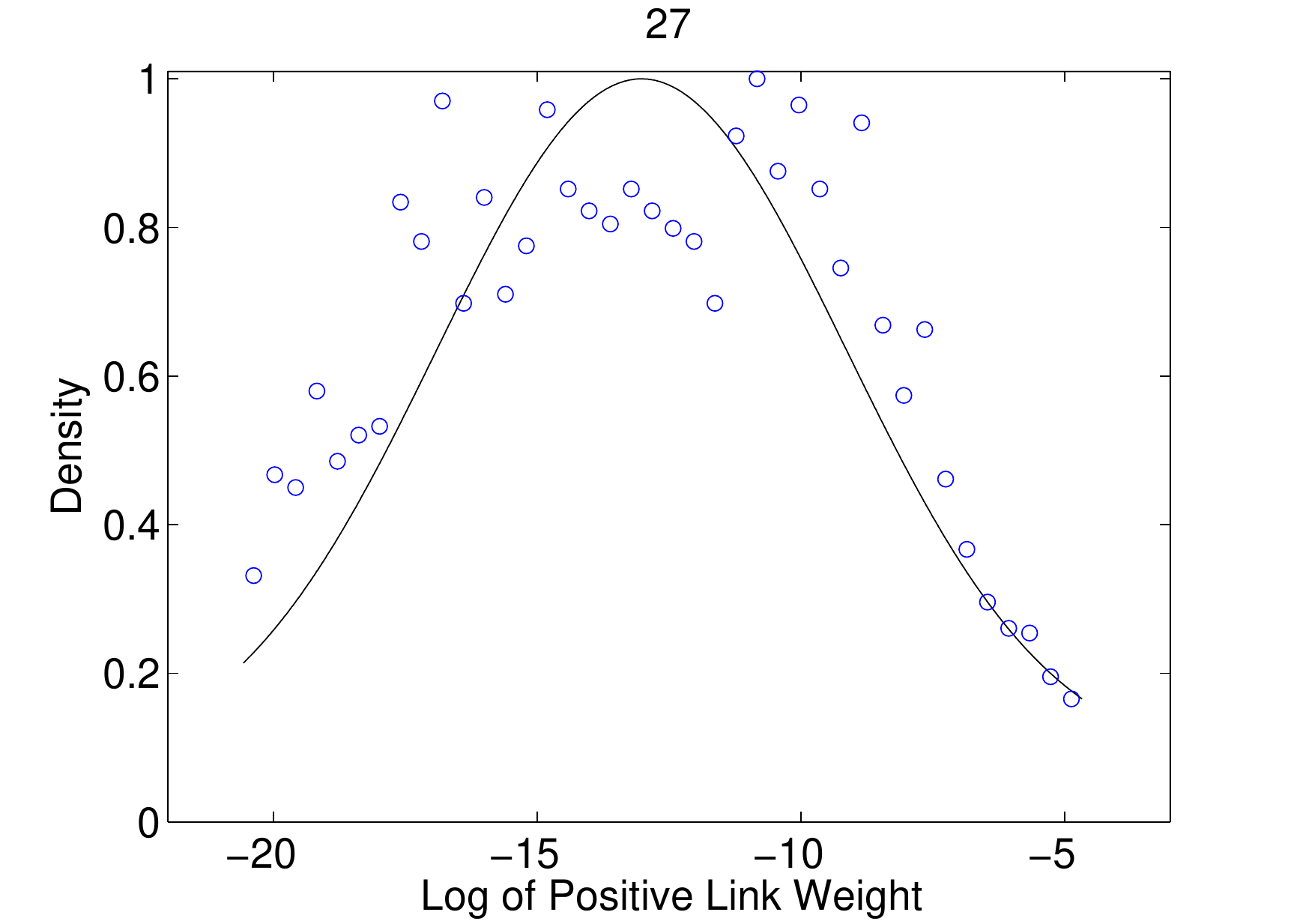}
\end{minipage}
\begin{minipage}{7cm}
\includegraphics[width=7cm]{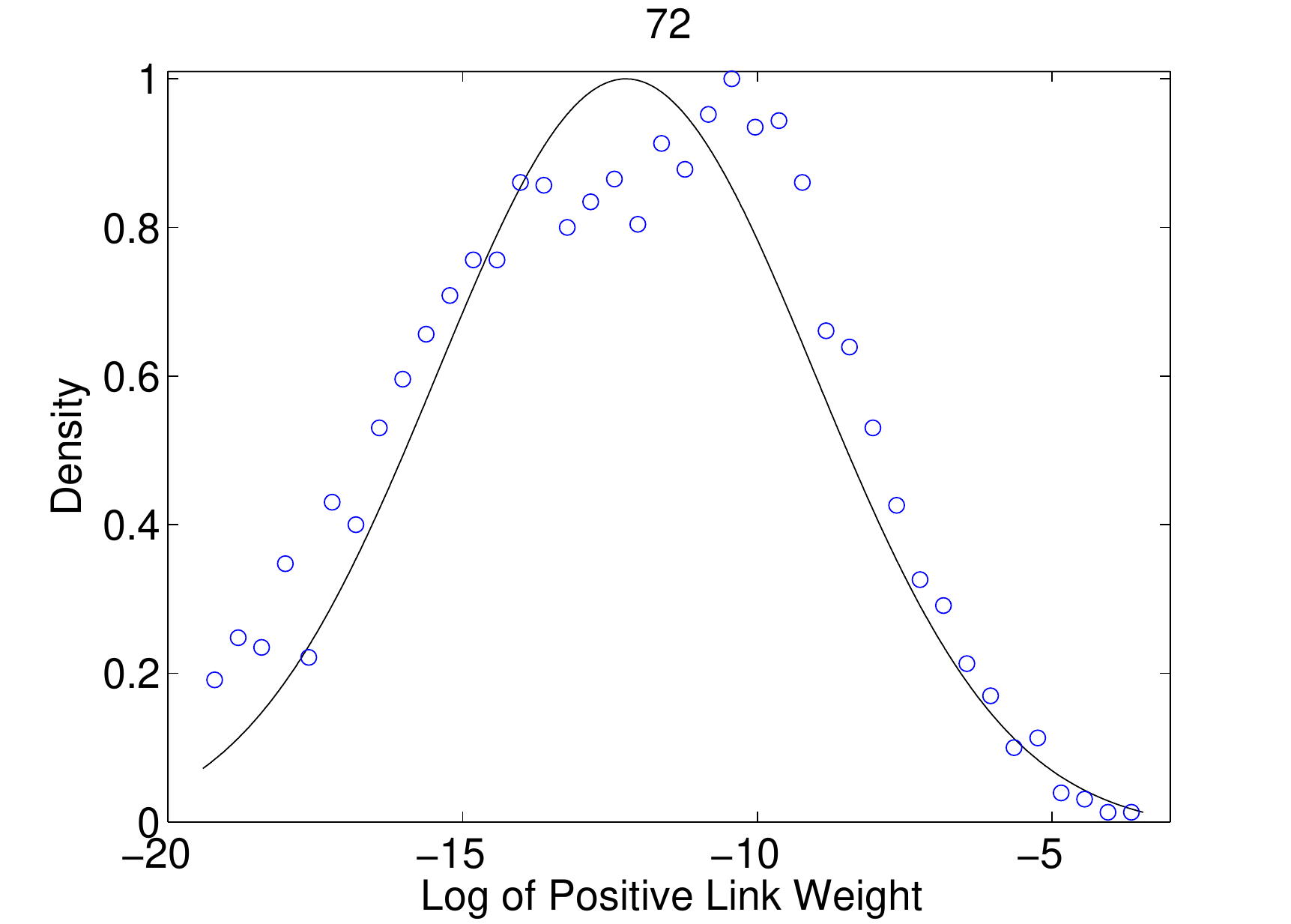}
\end{minipage}
\begin{minipage}{7cm}
\includegraphics[width=7cm]{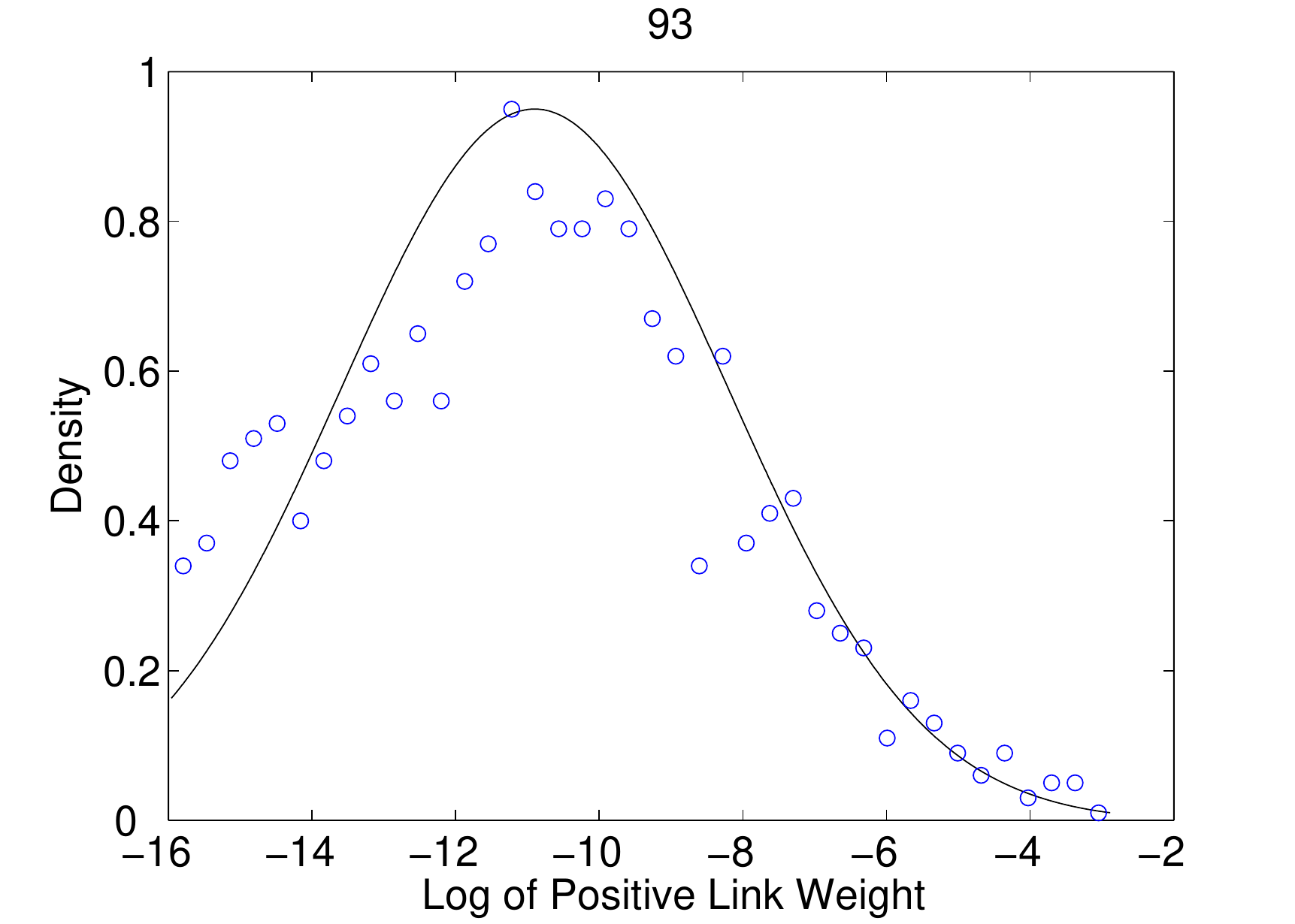}
\end{minipage}
\caption{{\label{fig:linkweight_distr}}Distributions of positive link weights
in 2003. 10: Cereals; 27: Mineral Fuels; 72: Iron and steel; 93: Arms. Solid line: Normal fit.}
\end{center}
\end{figure}

To begin exploring the issue whether log-normal aggregate distributions are the
result of heterogeneous, possibly non log-normal, commodity-specific
distributions, we have run a series of goodness-of-fit exercises \footnote{To
test for equality of two distributions we have employed the two-sample
Kolmogorov-Smirnov test. Testing for normality of logs of link weights has been
carried out using the Lilliefors and the one-sample Kolmogorov-Smirnov test
\citep{DagoSteph1986,Thode2002}.} to test whether: (i) any two pairs of
commodity-specific networks are characterized by the same link-weight
distribution; (ii) commodity-specific link-weight distributions are log-normal
(i.e., logs of their positive values are normal). Our result show that the body
of the aggregate distribution can be well-proxied by a log-normal, whereas the
upper tail seems to be thinner than what expected under log-normality (less
high-intensity links as expected). This means that log-normality found by
\citep{Fagiolo2009pre} may be also the outcome of symmetrization, i.e. of
studying a undirected weighted version of the ITN. We also find that only in
4\% of all the possible pairs of distributions ($4656=97*96/2$), the p-value of
the associated two-sided Kolmogorov test is greater than 5\%. These implies
that link-weight distributions are extremely heterogeneous across commodities.
Furthermore, according to both Lilliefors and one-sample normality Kolmogorov
tests, the majority of distributions seem to be far from log-normal densities,
see Figure \ref{fig:linkweight_distr} for some examples. This suggests that the
outcome of quasi-log-normality of link weights of the overall network may be a
sheer outcome of aggregation.

\subsection{Connected Components} \label{SubSec:ConnComps}
We now turn to analyzing the connectivity patterns of the binary aggregate and
commodity-specific trade networks by studying the size and composition of their
largest connected components.

If we employ the weaker definition of connectivity between two nodes
in a directed graph (either an inward or an outward link in place),
then the aggregate ITN is fully connected, i.e. the largest
connected component (LCC) contains all $N$ countries. If we instead
use the stronger definition (both the inward or the outward link in
place), then the aggregate network is never completely connected in
the time interval under analysis, and the composition of the LCC
changes with time. Table \ref{Tab:ConnCompAggr} shows the percentage
size of the LCC for the aggregate network, disaggregated according
to geographical macro-areas (i.e., we only consider the LCC in the
sub-network of the aggregate ITN made only of countries belonging to
any given geographical macro area). In Europe trade links are almost
always reciprocated and we notice the fast integration of Eastern
Europe after the mid 90s. Sub-Saharan Africa is the area where we
find the majority of countries without bilateral trade with other
countries of the area, a sign of poor trade connectivity perhaps
related to wars, trade barriers, lack of infrastructures, etc..

It is interesting to compare the above considerations about the reciprocity
structure of the international trade network  with a series of results
\cite{Garla2004,Garla2005} performed on a different dataset reporting aggregate
trade over the longer period 1950-2000 \cite{GledData2002}. Those analyses
reveal that the reciprocity has been fluctuating about an approximately
constant value up to the early 80�s, and has then been increasing steadily. In
other words, the international trade system appears to have undergone a rapid
reciprocation process starting from the 80�s. At the same time, the fraction of
pairs of countries trading in any direction (i.e. the density of the network
when all links are regarded as undirected) displays a constant trend over the
same period. Therefore, while at an undirected level there is no increase of
link density, at a directed level there is a  steep increase of reciprocity.
The combination of these results signals many new directed links being placed
between countries that had already been trading in the opposite direction,
rather than new pairs of reciprocal links being placed between previously
noninteracting countries. Thus, at an aggregate level many pairs of countries
that had previously been trading only in a single direction have been
establishing also a reverse trade channel, and this effect dominates on the
formation of new bidirectional relationships between previously non-trading
countries.

We turn now to analyze connectivity patterns of commodity-specific networks. In
this case, it is more reasonable to assume that two countries are connected in
a given commodity-specific network if they are linked either by an import or
export relationship (the weaker assumption above). Unlike the aggregate
network, no commodity-specific graph is completely connected. In what follows,
for the sake of exposition, we focus on year 2003 and we report connectivity
results for our 14 top commodities. Table \ref{Tab:ConnCompDisaggr} reports the
size of the LCC in different setups as far as the threshold $\underline{w}_t^c$
for the determination of binary relationships is concerned
($\underline{w}_t^c=0$, $\underline{w}_t^c=w_t^{c,p}$, where $w_t^{c,p}$ is the
p-th percentile of the link-weight distribution, with $p=90\%,95\%,99\%$). When
all trade fluxes are considered in the determination of a binary link, then all
commodity-specific networks are highly connected, and the size of the LCC is
relatively close to network size (except for the case of arms). If one raises
the lower threshold and only considers the 10\%, 5\% and 1\% strongest link
weights in each matrix, then few countries remain connected. For each
commodity, Table \ref{Tab:ConnCompComposition} lists the countries belonging to
the LCC in year 2003 and for the strongest 1\% links. It is easy to see that
the ``usual suspects'' (USA, Germany, Japan, etc.) belong to almost all
commodity LCCs. Some of them are unexpectedly small (coffee, cereals), others
are very large even if one is only focusing on a few largest trade links. All
in all, this evidence indicates that complete connectivity in the ITN is mainly
achieved through weak links, whereas strong links account for tightly
interconnected clubs that trade with each other not only in the aggregate but
also every possible commodity.

\subsection{Country rankings} \label{SubSec:Rankings}
In this subsection we analyze country rankings in 2003 according to the
alternative topological properties studied in the paper. For each node
statistic, we rank in a decreasing order countries in the panel and we report
the top-3 positions for our 14 benchmark commodities, as well as for the
aggregate network. Results are in Tables
\ref{Tab:RankNS}--\ref{Tab:RankWCC_CENTR}.

As far as node strength is concerned, USA, Germany, China and UK exhibit top
values of both import shares and output shares in almost all commodity
networks. These are the countries that trade more irrespective of the specific
commodity. Russia, Saudi Arabia and Norway top the fuel export ranking, Brazil
excels in coffee export, whereas Hong Kong and Mexico enter the top-3 positions
in cotton and cereals, respectively. ANNS rankings (Table \ref{Tab:RankANNS})
are more instructive, because they reveal that countries trading with partners
that imports/exports more, are typically small economies located outside Europe
and North America. This points to a general disassortative structure of the
network also at the commodity-specific level, a structural pattern that has
been observed in the aggregate as well in previous studies
\citep{Garla2004,Fagiolo2008physa}.

Rankings of clustering, on the other hand, display a markedly-larger commodity
heterogeneity in terms of countries appearing in the top-3 positions. Table
\ref{Tab:RankWCC_CENTR} shows results about overall weighted clustering, i.e.
the relative intensity of trade triangles with the target country as a vertex,
irrespective of the direction of trade flows. Notice that in the aggregate USA,
Germany and China are the most clustered nodes, but they do not always show up
in the same positions in all commodity rankings. This means that they typically
form extremely strong triangles in a few commodity networks (e.g., for USA
pharmaceutical, optical instruments). Note also the high-clustering levels
reached by Colombia in coffee trade, Algeria in cereals, Equatorial Guinea in
mineral fuels and organic chemicals, Uzbekistan in cotton. These are countries
that tend to be involved with a relevant intensity only in one particular type
of trade triangle, e.g. in-type for Algeria, out-type for Equatorial Guinea,
Uzbekistan and Colombia. This suggests, for example, that Algeria is very
likely to import cereals from two countries that are also trading cereals very
much. Similarly, Equatorial Guinea, Uzbekistan and Colombia tend to intensively
export mineral fuels, cotton and coffee, respectively, to pairs of countries
that also trade intensively these commodities together. Finally, centrality
rankings shed some light on the relative positional importance of countries in
the network. Rankings stress, beside the usual list of large and influential
countries, the key role played by Switzerland in precious metals, Russia, Saudi
Arabia and Norway in mineral fuels, Indonesia in coffee and Thailand in
cereals.

\subsection{Correlations between topological properties within commodity networks} \label{SubSec:CorrAcross}
Early work on the aggregate ITN has singled out robust evidence about the
correlation structure between topological properties
\citep{Garla2004,Garla2005,Garla2007,Fagiolo2008physa,Fagiolo2009pre,Fagiolo2009jee}.
For example, disassortative patterns (negative correlation between ANND/ND and
ANNS/NS; see also above) has been shown to characterize the binary ITN
(strongly) and the weighted ITN (weakly). Also, the aggregate ITN exhibits a
trade structure where countries that trade more intensively are more clustered
and central. Here we check whether such structure is robust to disaggregation
at the commodity level by comparing the correlation between different
topological properties (e.g., $NS_{in}$ vs. $ND_{in}$) within each commodity
network. In the next section, conversely, we shall look at how the same
topological property (e.g., $NS_{in}$) correlates across different networks.

Table \ref{Tab:CorrWithin} shows the most interesting correlation coefficients
between node statistics \footnote{More precisely, the correlation coefficient
between two node statistics related to the same commodity network $c$, i.e.
$X^c$ and $Y^c$, is defined here as the product-moment (Spearman) sample
correlation, i.e.
$\sum_{i}{(x_i^c-\overline{x^c})(y_i^c-\overline{y^c})}/[(N-1)s_X^c s_Y^c]$,
where $\overline{x^c}$ and $\overline{y^c}$ are sample averages and $s_X^c$ and
$s_Y^c$ are sample standard deviations across nodes in network $c$.}. Note
first that, all in all, the sign of any given correlation coefficient computed
for the aggregate network remains the same across almost all commodity-specific
networks. This is an interesting robustness property, as we have shown that
commodity-specific networks are relatively heterogeneous according to e.g. the
shape of their link-weight distribution. It appears instead that despite
heterogeneously-distributed link weights the inherent architecture of
commodity-specific networks mimics those of the aggregate (or viceversa).

Almost all the signs are in line with what previously observed. For example,
countries that trade with more partners also trade more intensively (both as
exporters ad importers). Furthermore, countries that import (export) more,
typically import from (export to) countries that in turn export on average
relatively less (disassortativity). The magnitude of this disassortativity
pattern is however different according to whether one looks at imports of
exports. On average, countries that import from a given country, trade
relatively less than those that export to the same country, i.e. the magnitude
of the correlation coefficients between $NS_{out}$ and both $ANNS_{out-in}$ and
$ANNS_{out-out}$ is larger than the magnitude of the correlation coefficients
between $NS_{in}$ and both $ANNS_{in-in}$ and $ANNS_{in-out}$.

Another robust correlation pattern that emerges is about clustering and
centrality. Countries that trade more in terms of their node strengths are also
more clustered and more central. This happens irrespectively of the commodity
traded.

The only partial exceptions to such evidence are represented by the commodity
networks of cereals and mineral fuels. For example, countries that imports
relatively more cereals (mineral fuels) typically import from countries that
also export (import) more cereals (mineral fuels). This does not happen however
for exports of such commodities, as correlations are negative or very close to
zero. Also, countries that trade more these two commodities are relatively less
clustered than happens in other commodity classes.

\subsection{Correlations between topological properties across commodity networks} \label{SubSec:CorrWithin}
In the latter subsection we have investigated correlations computed between
different node topology statistics within the same network. We now explore
correlation patterns of node statistics across commodity networks. More
precisely, for each given node statistic $X$, we compute all possible
$C(C-1)/2=4656$ correlation coefficients:

\beq
\rho^{c,c'}(X)=\frac{\sum_{i=1}^{N}{(x_i^c-\overline{x}^c)(x_i^{c'}-\overline{x}^{c'})}}{(N-1)s_X^c
s_X^{c'}}, \eeq where $\overline{x}^c$ and $\overline{x}^{c'}$ are sample
averages and $s_X^{c}$ and $s_X^{c'}$ are sample standard deviations across
nodes in network $c$ and $c'$.

\begin{sidewaysfigure}[t]
\begin{center}
\begin{minipage}{7cm}
\includegraphics[width=7cm]{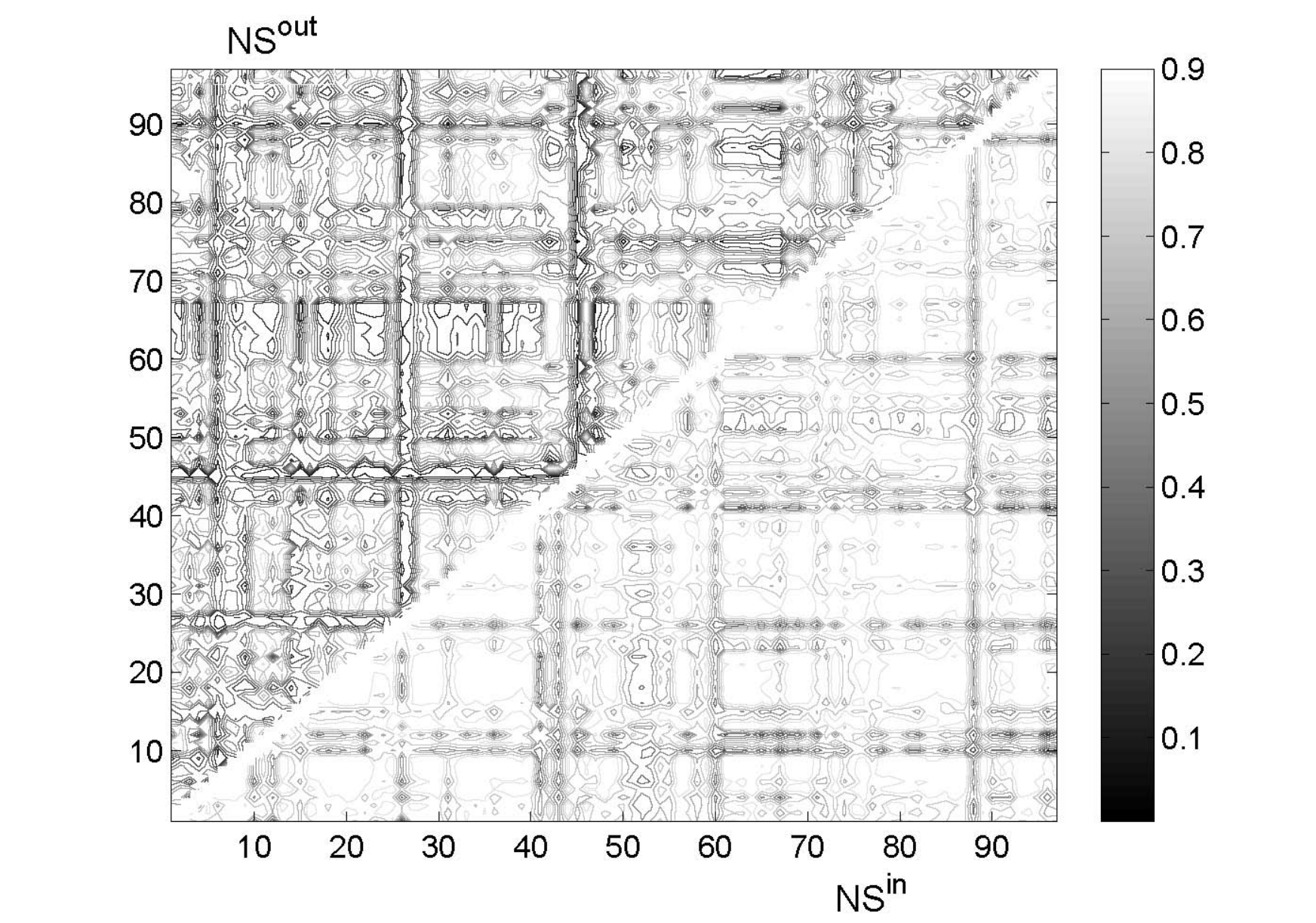}
\end{minipage}
\begin{minipage}{7cm}
\includegraphics[width=7cm]{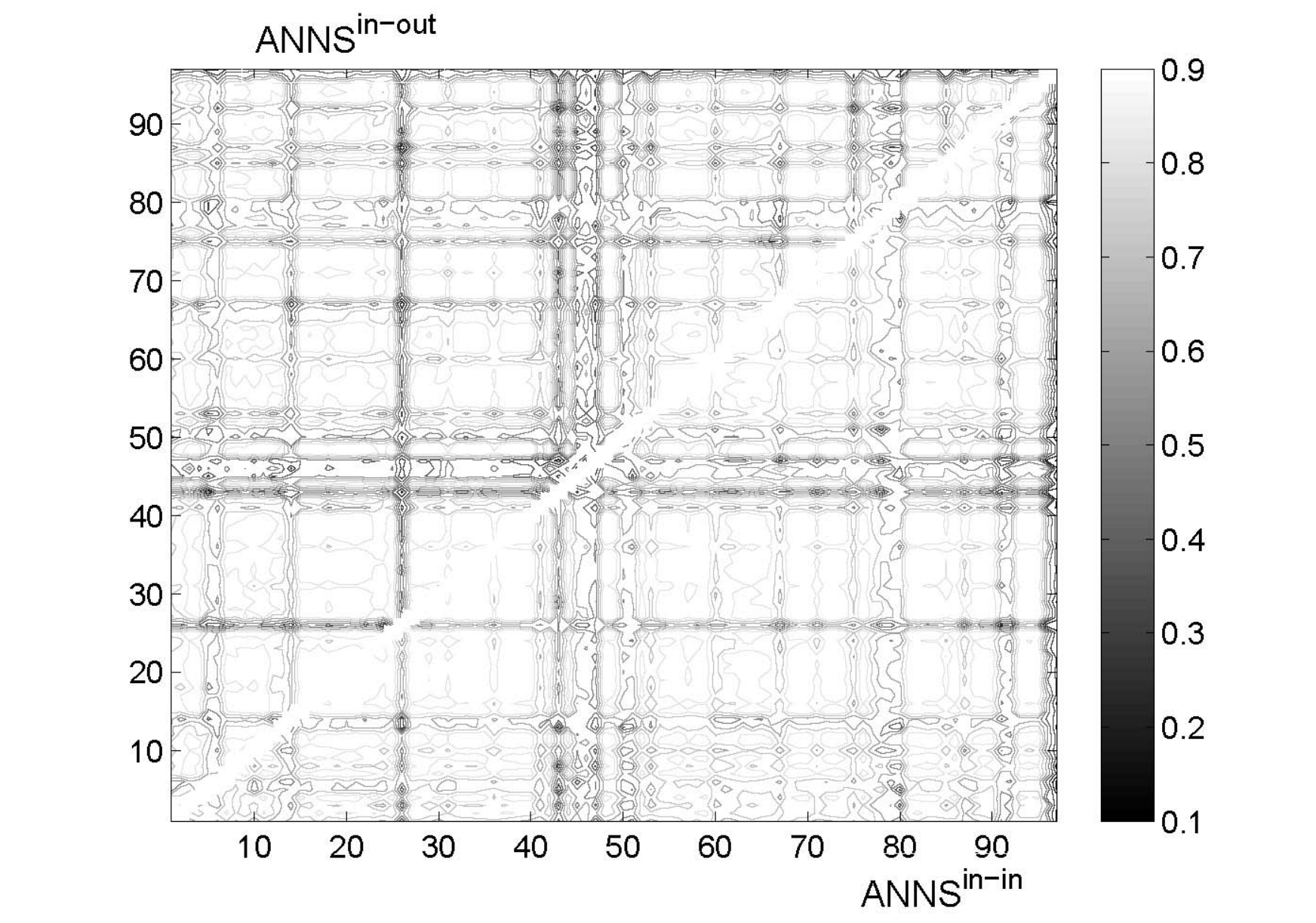}
\end{minipage}
\begin{minipage}{7cm}
\includegraphics[width=7cm]{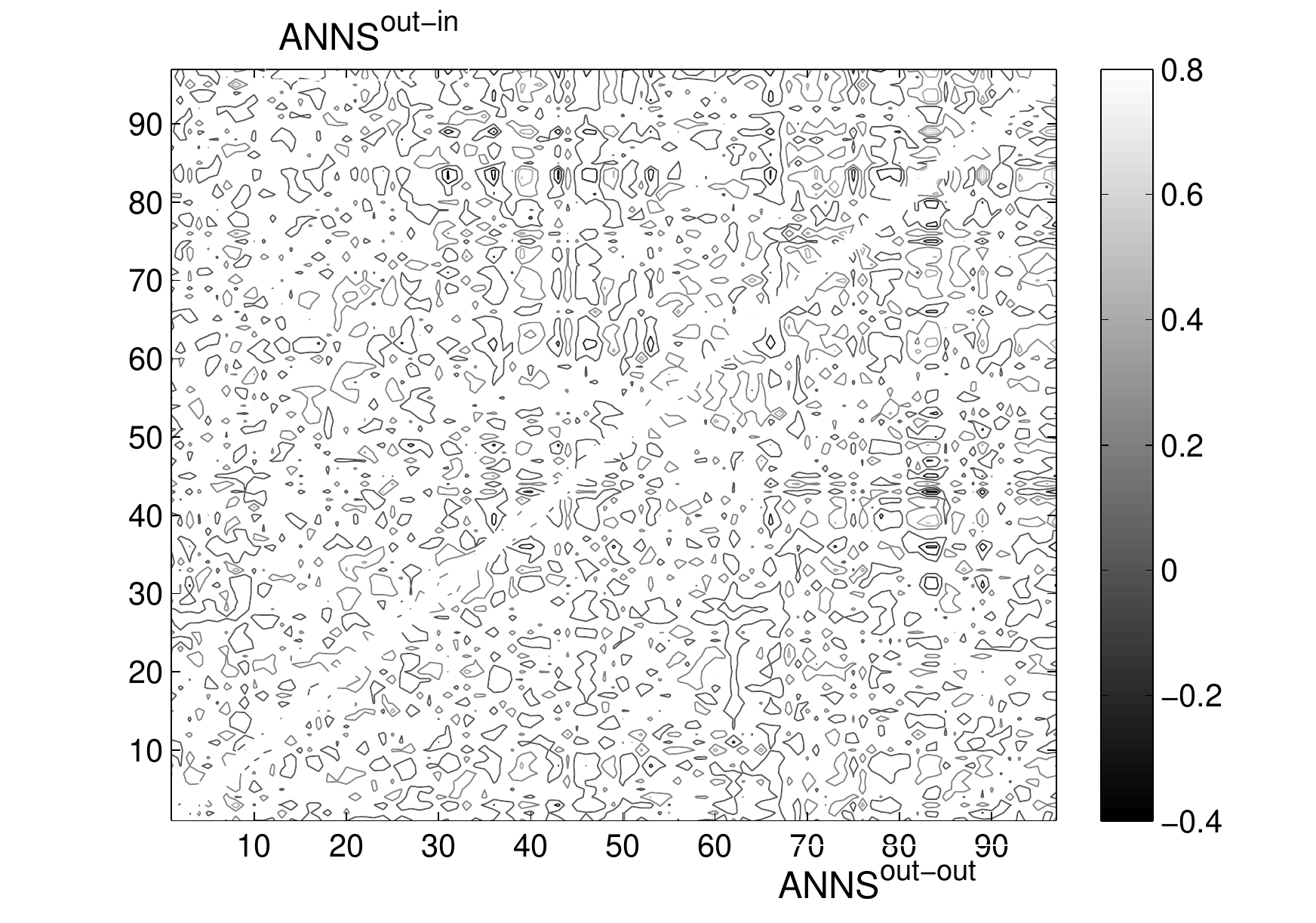}
\end{minipage}
\begin{minipage}{7cm}
\includegraphics[width=7cm]{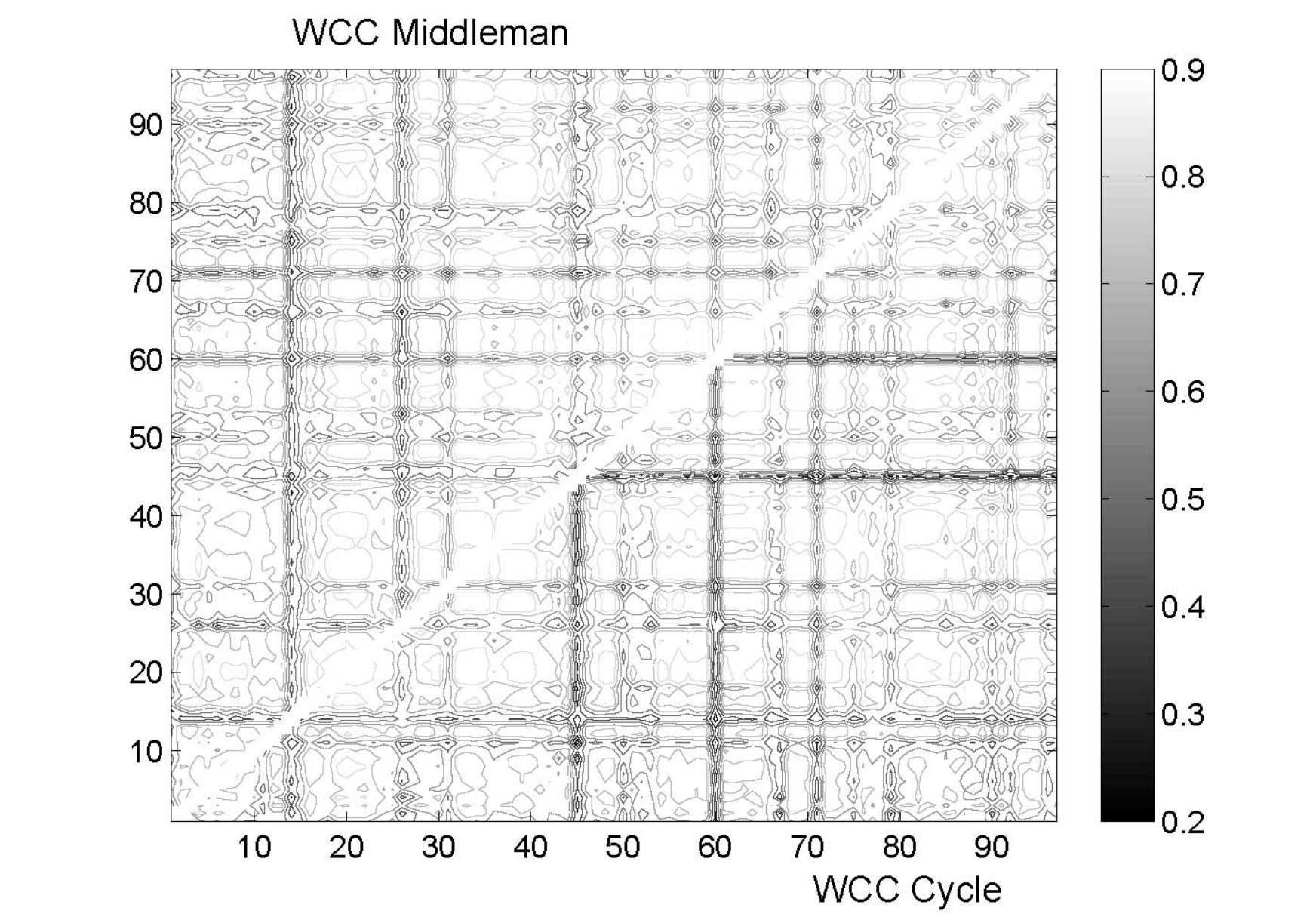}
\end{minipage}
\begin{minipage}{7cm}
\includegraphics[width=7cm]{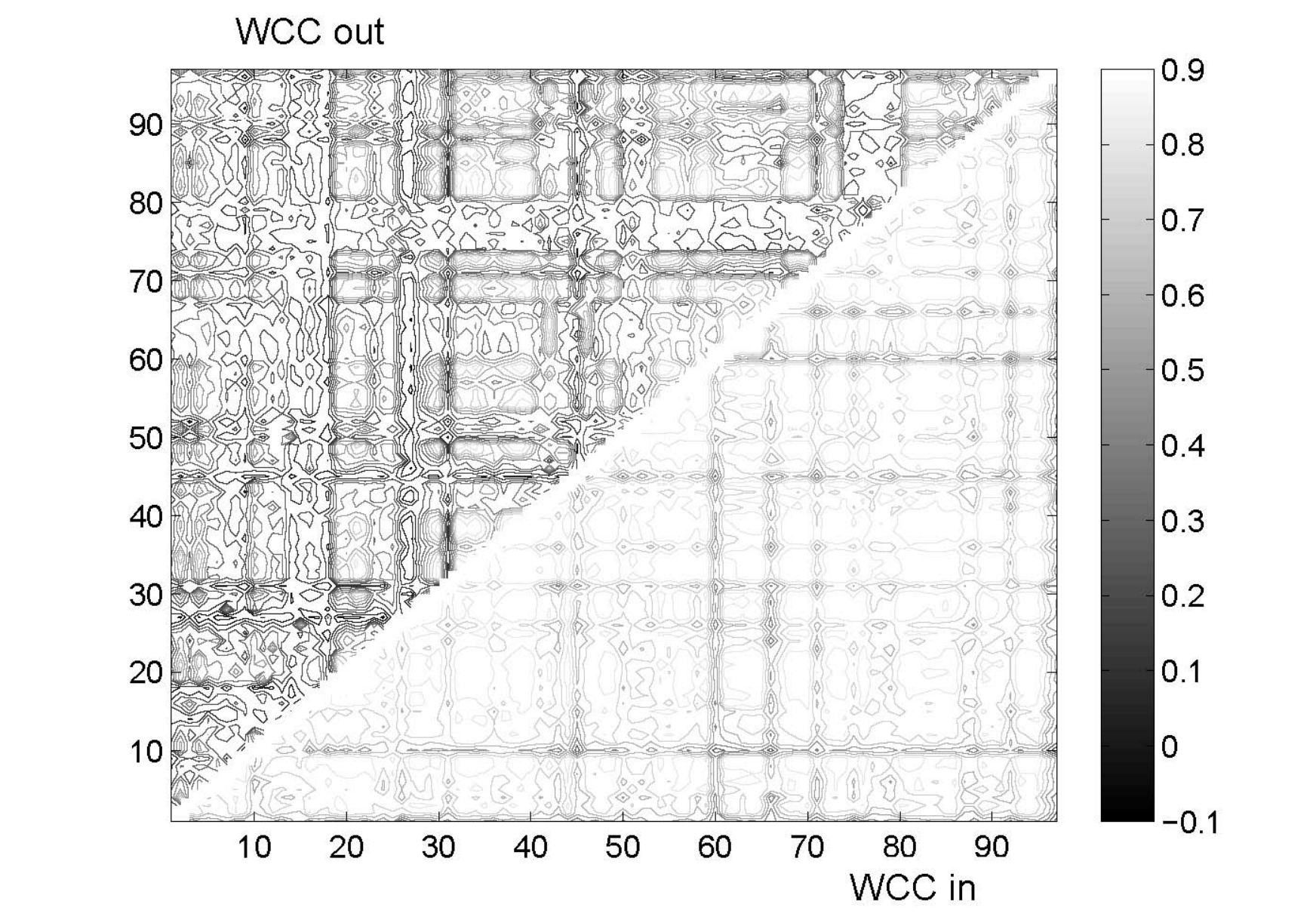}
\end{minipage}
\caption{{\label{fig:corr_cross}}Correlation coefficients of topological
statistics across networks. Axes represent HS classification codes. When
convenient, each plot contains the correlation patterns for two statistics, one
in the upper-left triangle and another in the bottom-right triangle.}
\end{center}
\end{sidewaysfigure}

Figure \ref{fig:corr_cross} plots correlation patterns for some node statistics
\footnote{Since correlations are symmetric, each figure actually reports
---when convenient--- correlations for two statistics, one in the upper-left
triangle and the other in the lower-right triangle. Axes stand for HS codes.}.
Notice first that on average correlation coefficients are always positive for
both $NS_{in}$ and $NS_{out}$, but those for $NS_{in}$ are larger than those
for $NS_{out}$. This suggests that in general if a country exports (imports)
more of a commodity, then it exports (imports) more of all other commodities.
However, imports of different commodities are much more correlated than
exports. This may be intuitively explained by the fact that (according to the
HS classification) country imports may be related to inputs in the production
process, which requires many different commodities. Instead, exports mainly
regards the output process and they might therefore depend on the patterns of
specialization of a country. The same behavior characterizes in- and out-types
of clustering: countries that form intensive triangles where they import from
two intensively-trading partners do so irrespectively of the commodity traded,
but the correlation is higher than the corresponding pattern when now countries
exports two intensively-trading partners.

An additional interesting insight comes from observing that in many cases
darker stripes and lighter squares characterize the plots. Darker stripes are
located typically on the edge between two adjacent 1-digit commodity classes,
whereas squares with similar shades cover the entire 1-digit class. This means
that in general correlation patterns mimic the HS classification, i.e.
across-network correlations of a given statistics look similar when the
commodity is similar according to the HS class ---or abruptly change when one
moves from a commodity class to another representing structurally different
products and services. Interestingly, darker stripes often correspond to
commodities that are less likely to be used as inputs then produced as outputs
(manufactured product, typically retail oriented).

The fact that their statistics are more weakly correlated with those of other
commodities hints to two different patterns as far as imports/exports and
specialization patterns are concerned, and calls for further and deeper
analyses. The fact that results partly mimic (or depend from) the
classification scheme used indicate that it would be interesting to find
classification-free grouping of commodities that are more data-driven. Data on cross-commodity correlations may be employed to address this issue, as we begin to study in the next section.
The method we propose to study the problem is general, and represents a first step towards a systematic approach to the analysis of large multi-networks.

\section{A framework for multi-network analysis\label{Sec:sec_corr}}
The above results show that the international trade network is not simply a superposition of independent commodity-specific layers. We found that significant correlations among layers make a comprehensive understanding of the structural properties of the whole system challenging. In particular, while single layers can certainly be studied independently using standard tools of network theory, a novel and more general framework of analysis is required in order to consistently take into account how different networks interact with each other to form the emerging aggregated network.

This problem is general, and not restricted to the particular system we are considering here.
Besides a number of other economic and financial networks, that are virtually always systematically characterised by a superposition of product- or sector-specific relationships, other important examples include large social networks.
Real social webs are believed to be the result of different means of interaction among actors, with ties of different types (friendship, coaffiliation, relatedness, etc.) cooperating to create a multiplex social network. Traditionally however, experimental constraints have limited the availability of real data, especially if reporting the different nature of social ties, to small networks. More recently, with the increasing availability of detailed large social network data, disentangling the different types of social relations is becoming possible also at a larger scale. Thus the type of problem we are facing here is likely to become of common interest in the near future for many research fields.

In what follows we make a first step in this direction by proposing a simple approach to characterise the mutual dependencies among layers in multi-networks, and their hierarchical organization. This approach is simple and general, and can therefore prove useful in the future for the analysis of other multi-networks emerging as the interaction of different sub-networks.

\subsection{Interdependency of layers}
As a starting observation we note that, when studying a multi-network, the most detailed level of analysis focuses on the correlations between the presence, and the intensity in the weighted case, of single edges across different sub-networks.
Inter-layer correlations between more aggregated properties (such as those we showed above between commodity-specific node degrees, node strengths, clustering coefficients, etc.) are ultimately due to these fundamental edge-level correlations.
For this reason, one can perform a more detailed analysis by measuring inter-layer correlations according to any single observed interaction involving different layers.
This analysis is possible at both weighted and unweighted levels for all the $C(C-1)/2$ pairs of layers, where $C$ is
the total number of layers. As we show later on, the analysis of inter-layer correlations allows to define a hierarchy of layers.
In the particular case of the trade system, this results in a taxonomy of commodities according to their roles in the world economy. We note that recent studies have already focused on the analysis of similarities among commodities, and on
the associated reconstruction of a commodity space of goods, based on the
observed patterns of revealed comparative advantage for countries
\cite{Hidalgo_etal_2007_Science,Hidalgo_Hausmann_2009}, i.e. without
specifically considering the structure of trade flows across countries. By constrast, the method that we use here allows to make use of more detailed information.

To be explicit, for each pair of layers $(c,c')$, we consider the inter-layer
correlation coefficient $\phi^{c,c'}_w(t)$ between the corresponding edge
weights:
\begin{equation}\label{eq_phiw}
\phi^{c,c'}_w(t)\equiv\frac{\sum_{i\ne j}\left[w^{c}_{ij,t}-\overline{w^c_t}\right]\left[w^{c'}_{ij,t}-\overline{w^{c'}_t}\right]}
{\sqrt{\sum_{i\ne j}\left[w^{c}_{ij,t}-\overline{w^c_t}\right]^2
\sum_{i\ne j}\left[w^{c'}_{ij,t}-\overline{w^{c'}_t}\right]^2}},
\end{equation}
where the subscript $w$ indicates that we are explicitly taking into
account link weights, and $\overline{w^c_t}\equiv \sum_{i\ne
j}w^{c}_{ij,t}/N(N-1)$ is the weight of links embedded in layer $c$, averaged over directed pairs of vertices.
In our specific case study, $\overline{w^c_t}=1/N(N-1)$ is the traded volume of commodity
$c$ averaged across all directed pairs of countries, which is idependent of $c$ due to the choice of the normalization.
Similarly, if
one focuses only on the topology and discards weights, it is
possible to define the inter-layer correlation coefficient
\begin{equation}\label{eq_phiu}
\phi^{c,c'}_u(t)\equiv\frac{\sum_{i\ne j}[a^c_{ij}(t)-\bar{a}^c(t)][a^{c'}_{ij}(t)-\bar{a}^{c'}(t)]}
{\sqrt{\sum_{i\ne j}\left[a^c_{ij}(t)-\bar{a}^c(t)\right]^2
\sum_{i\ne j}[a^{c'}_{ij}(t)-\bar{a}^{c'}(t)]^2}}
\end{equation}
where $u$ stands for unweighted, and $\overline{a^c_t}\equiv \sum_{i\ne
j}a^{c}_{ij,t}/N(N-1)$ is the fraction, measured across all directed pairs of
vertices, of interactions involving layer $c$. Being Pearson's
correlation coefficients, $\phi^{c,c'}_w(t)$ and $\phi^{c,c'}_u(t)$ can take
values in the range $[-1,+1]$, the two extrema representing complete
anticorrelation and complete correlation respectively. Zero correlation is
expected for statistically independent, non-interacting layers. Note that both quantities already take
an overall size effect (total link weight and global link density respectively)
into account. Therefore they allow comparisons across different years even if
these overall properties are changing in time. For each year $t$ considered,
eq.(\ref{eq_phiw}) gives rise to a $C\times C$ \emph{weighted inter-layer
correlation matrix}
\begin{equation}\label{eq_matrixphiw}
\Phi_w(t)=\{\phi_w^{c,c'}(t)\}
\end{equation}
and eq.(\ref{eq_phiu}) gives rise to a $C\times C$ \emph{unweighted
inter-layer correlation matrix}
\begin{equation}\label{eq_matrixphiu}
\Phi_u(t)=\{\phi_u^{c,c'}(t)\}
\end{equation}
both matrices being symmetric and with unit values along the diagonal.

In the case considered here, the above matrices quantify on an empirical basis how correlated are edges belonging to
different layers. Large values of the correlation
coefficient $\phi^{c,c'}_u(t)$ signal that $c$ and $c'$ play similar roles in
the international trade system, as they are frequently traded together between
pairs of countries (i.e. they often share the same importer and exporter
country simultaneously). The quantity $\phi^{c,c'}_w(t)$ measures the same
effect, but also taking traded volumes into account. Although large
correlations should in principle be observed more frequently for commodities of
similar nature (``intrinsic'' correlations) as they are expected to be both
produced and consumed by similar sets of countries, they could be observed in
more general cases as well (``revealed'' correlations). Indeed, if
intrinsically different commodities turn out to be highly correlated this can
be interpreted as the result of favored trades of different goods between pairs
of countries. For instance, in case of common geographic borders, trade
agreements, or membership to the same free trade association or currency union,
two countries $i$ and $j$ may prefer to exchange various types of commodities
even if there are many potential alternative trade partners, either as
importers or as exporters, for each commodity. Conversely, inter-layer
correlations are decreased in presence of opposite trade preferences, i.e. by
the tendency of pairs of countries to have specialized exchanges involving
particular (sets of) commodities.

\begin{sidewaysfigure}[t]
\begin{center}
\begin{minipage}{7cm}
\includegraphics[width=7cm]{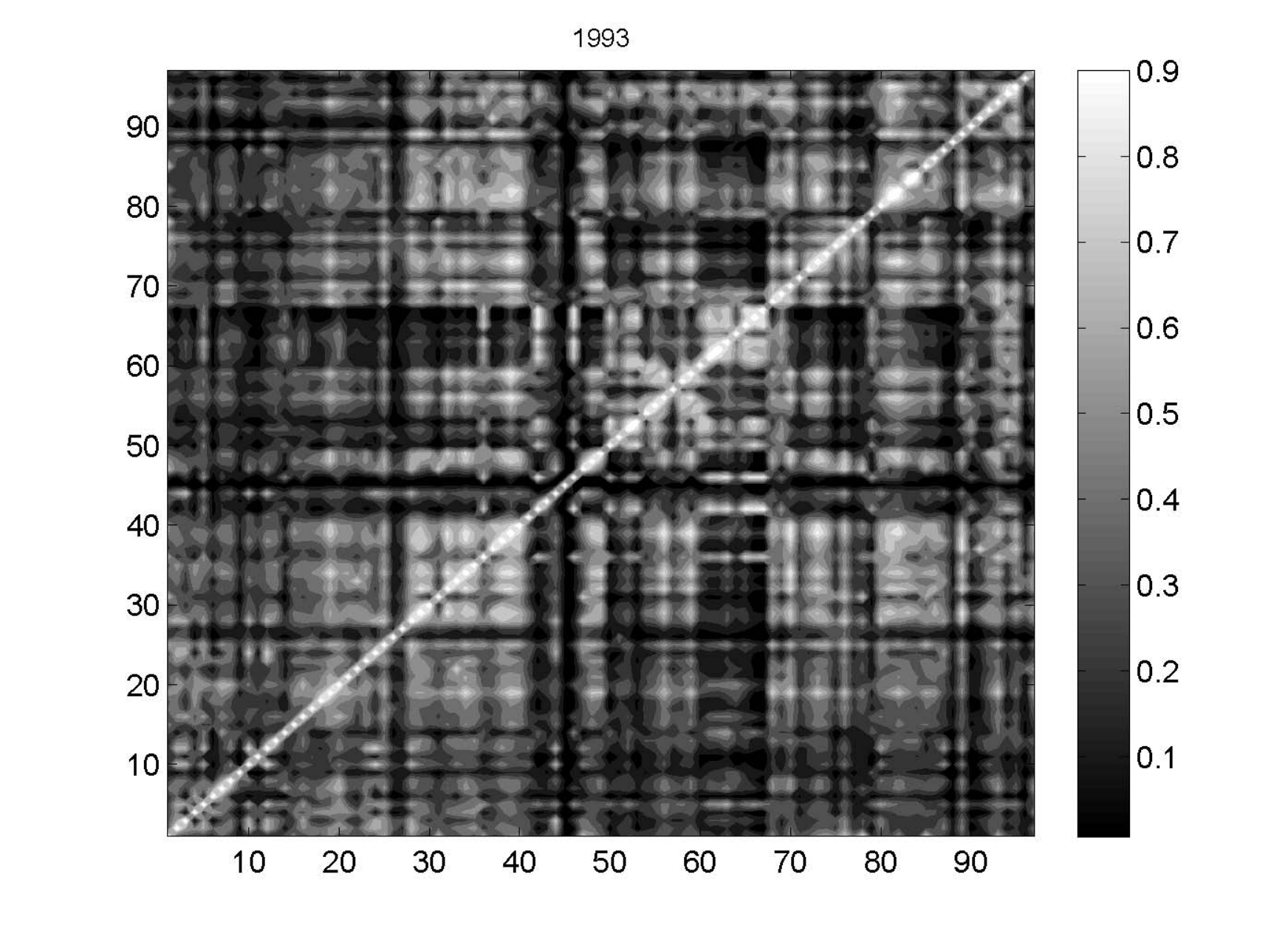}
\end{minipage}
\begin{minipage}{7cm}
\includegraphics[width=7cm]{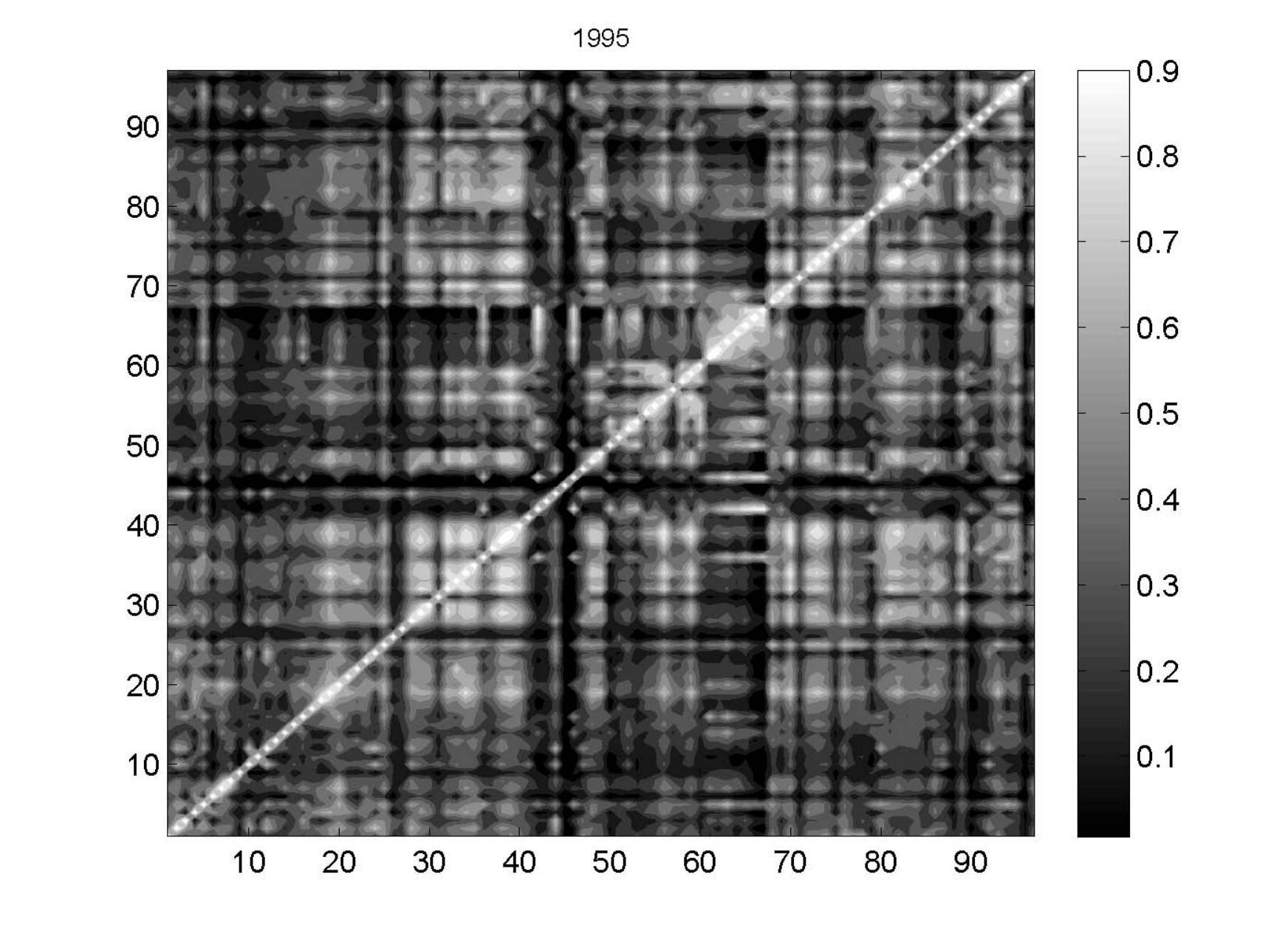}
\end{minipage}
\begin{minipage}{7cm}
\includegraphics[width=7cm]{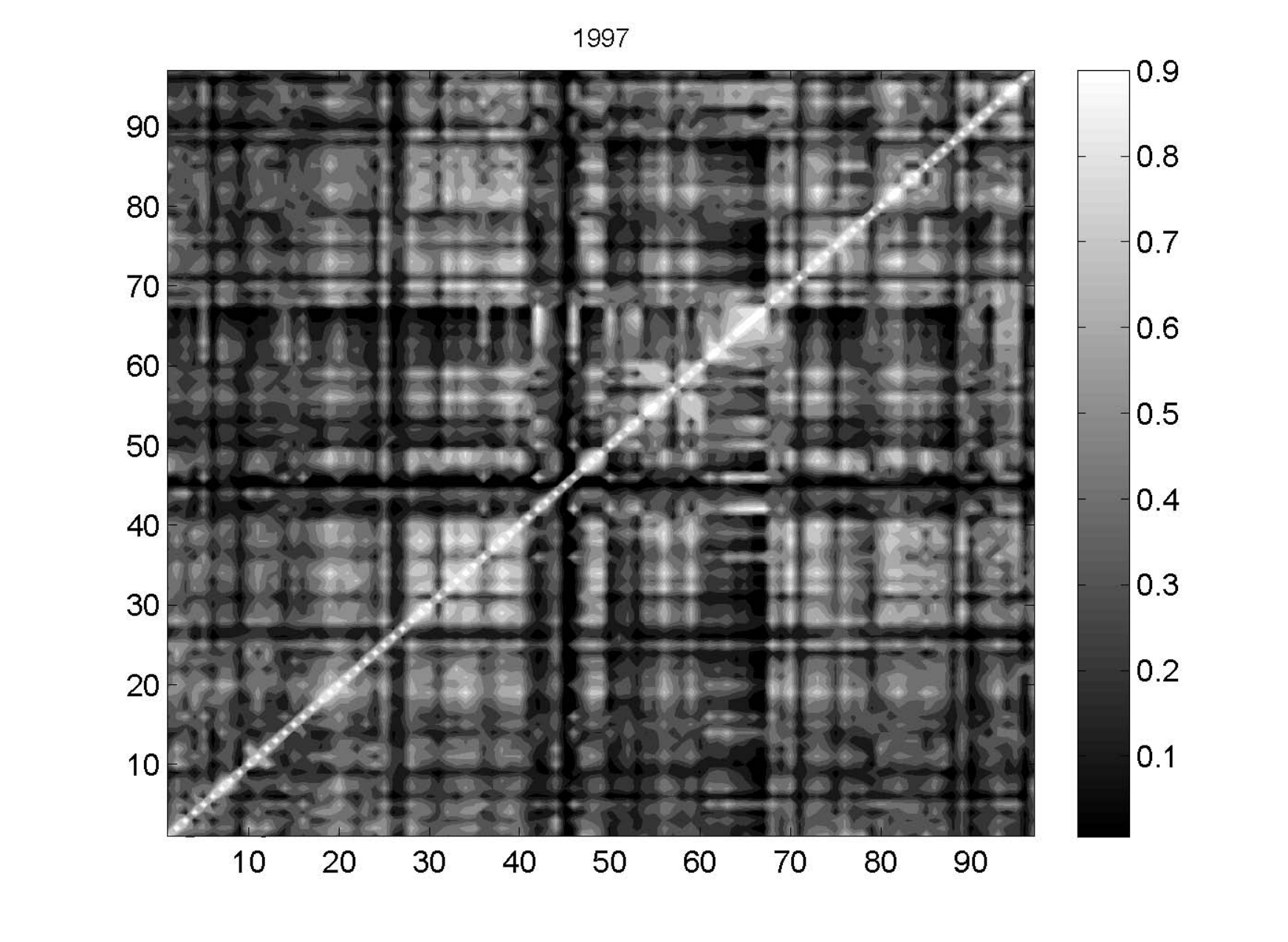}
\end{minipage}
\begin{minipage}{7cm}
\includegraphics[width=7cm]{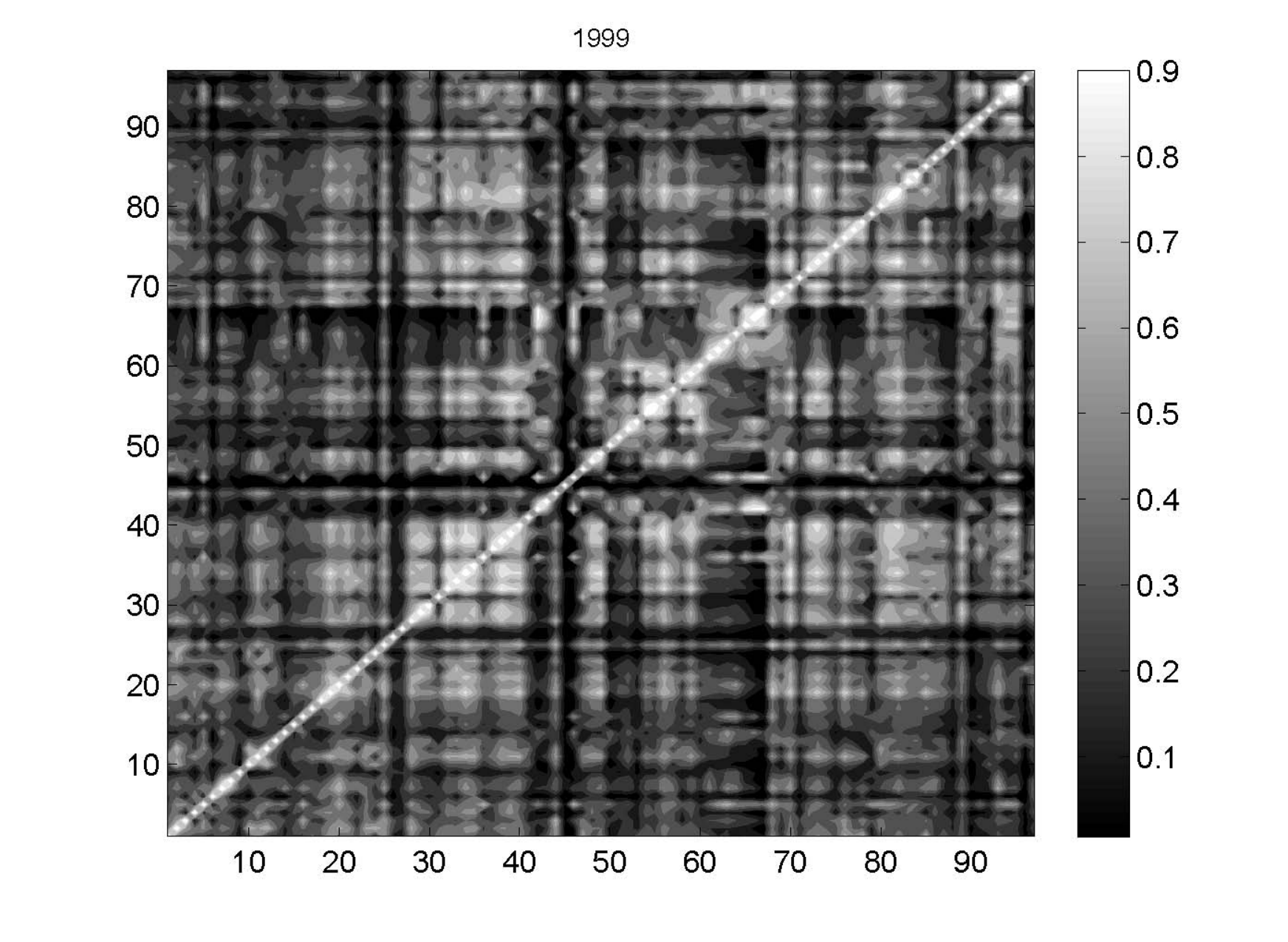}
\end{minipage}
\begin{minipage}{7cm}
\includegraphics[width=7cm]{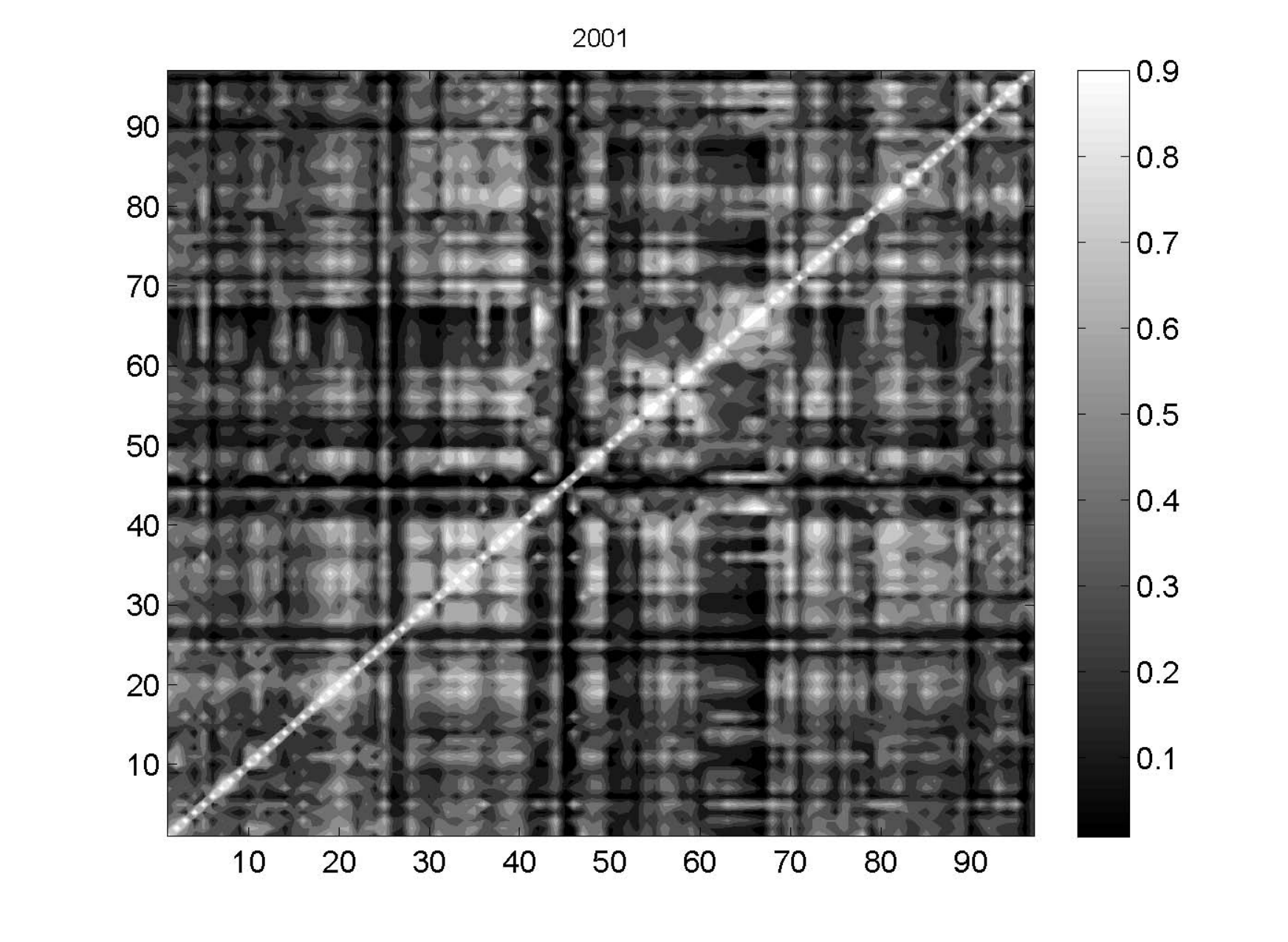}
\end{minipage}
\begin{minipage}{7cm}
\includegraphics[width=7cm]{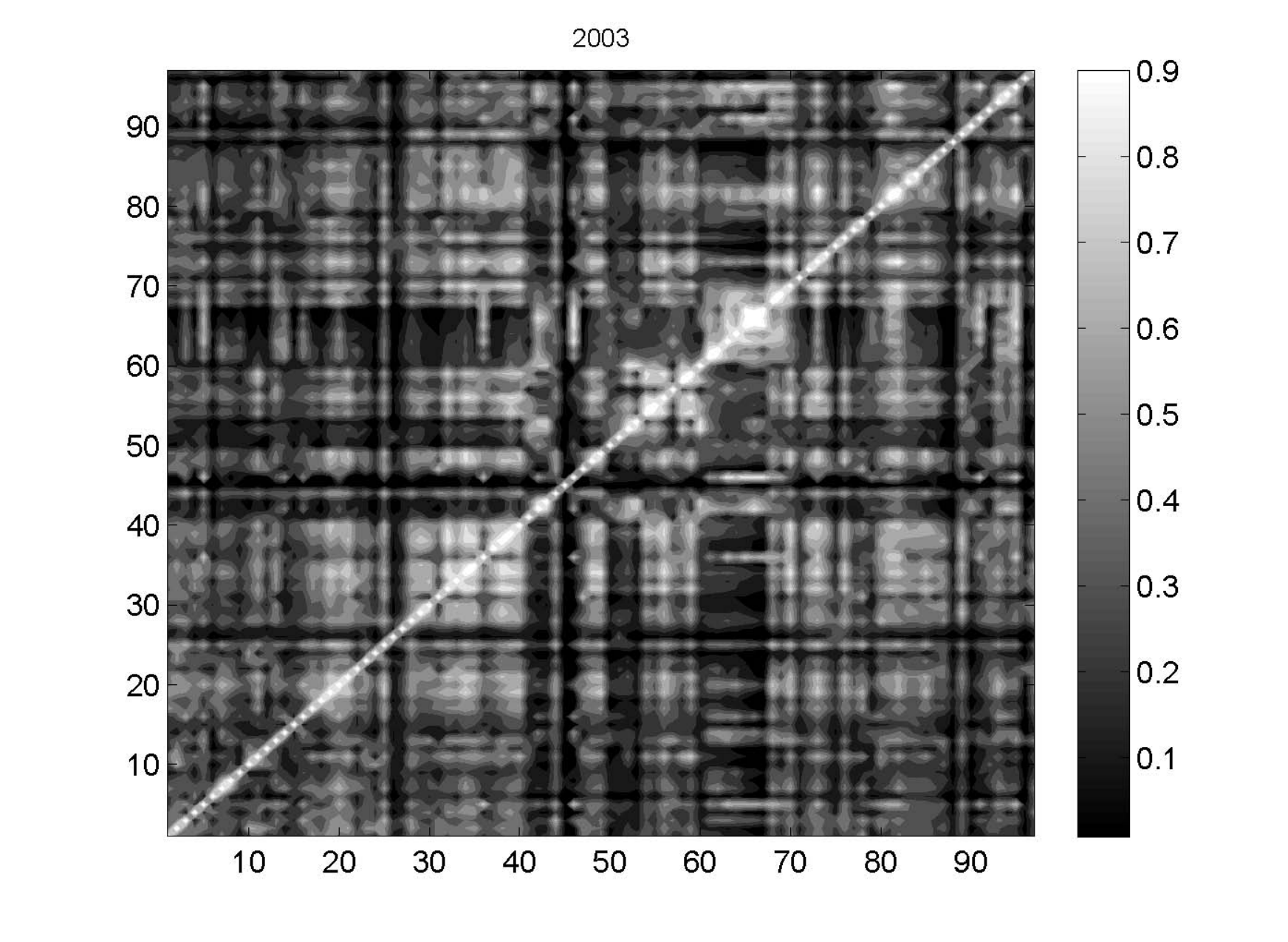}
\end{minipage}
\caption[]{\small Plots of weighted inter-layer correlation
matrices $\Phi_w(t)$ for years $t=1993,1995,1997,1999,2001,2003$.
\label{fig_phiw}}
\end{center}
\end{sidewaysfigure}

\begin{sidewaysfigure}[t]
\begin{center}
\begin{minipage}{7cm}
\includegraphics[width=7cm]{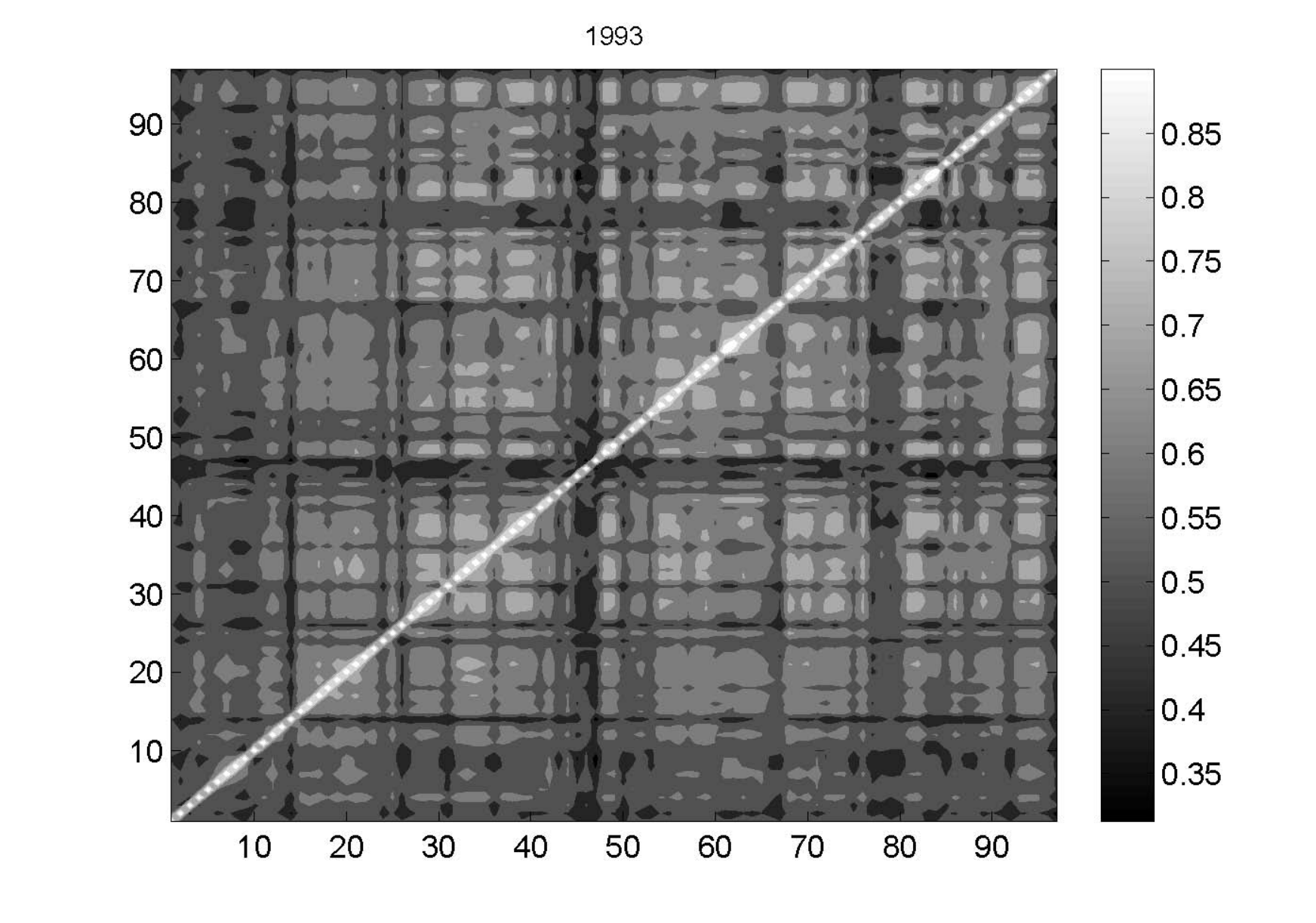}
\end{minipage}
\begin{minipage}{7cm}
\includegraphics[width=7cm]{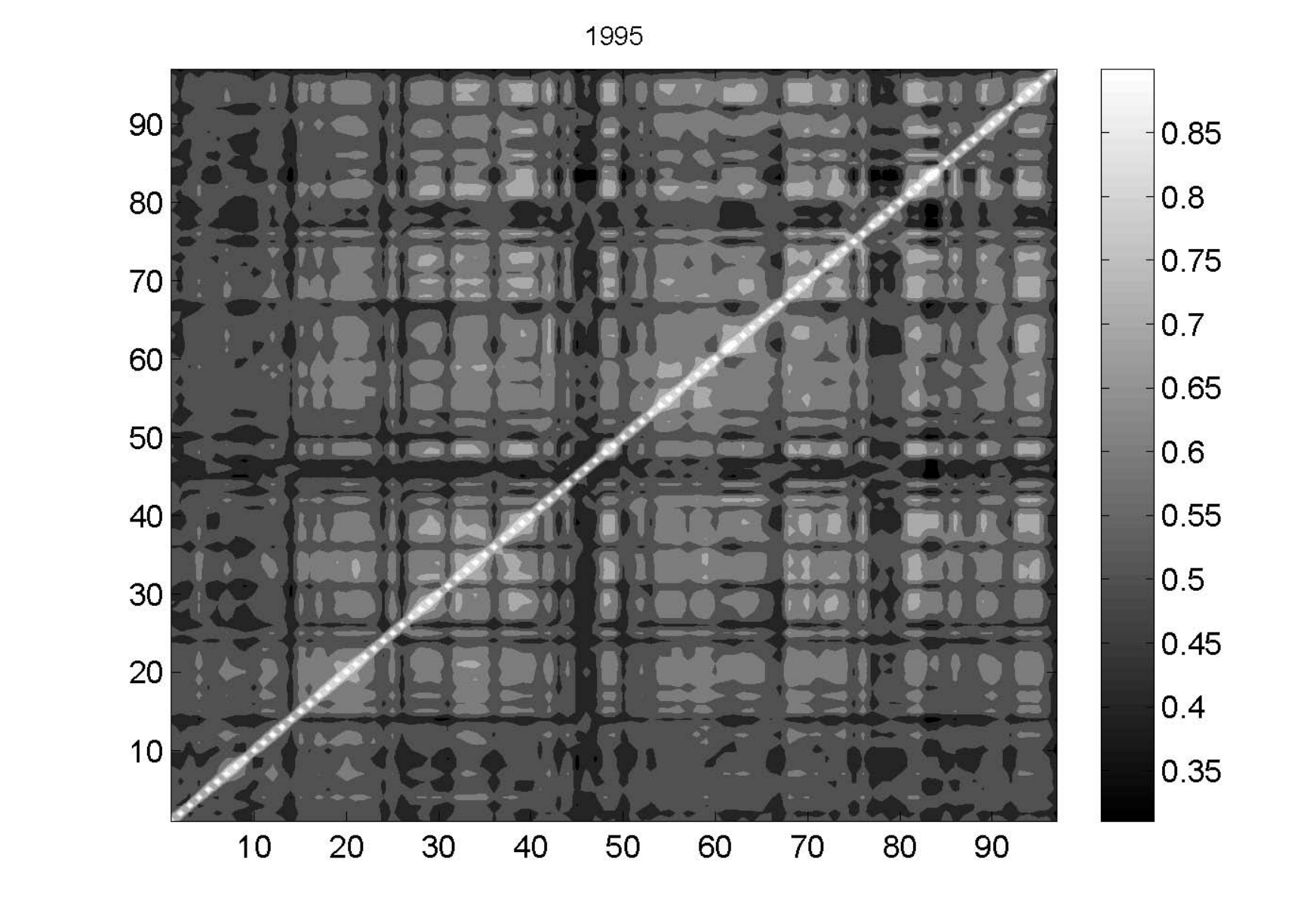}
\end{minipage}
\begin{minipage}{7cm}
\includegraphics[width=7cm]{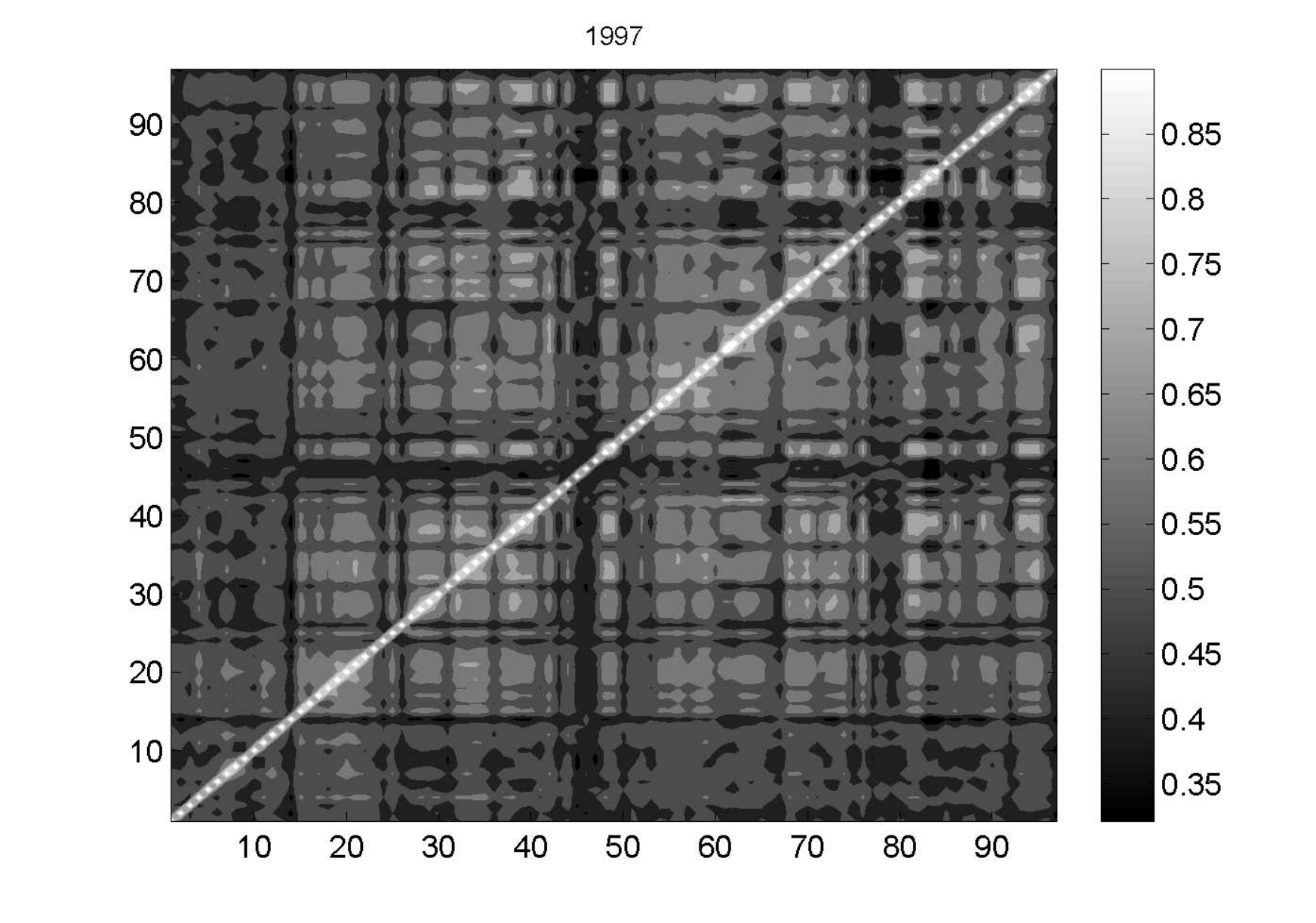}
\end{minipage}
\begin{minipage}{7cm}
\includegraphics[width=7cm]{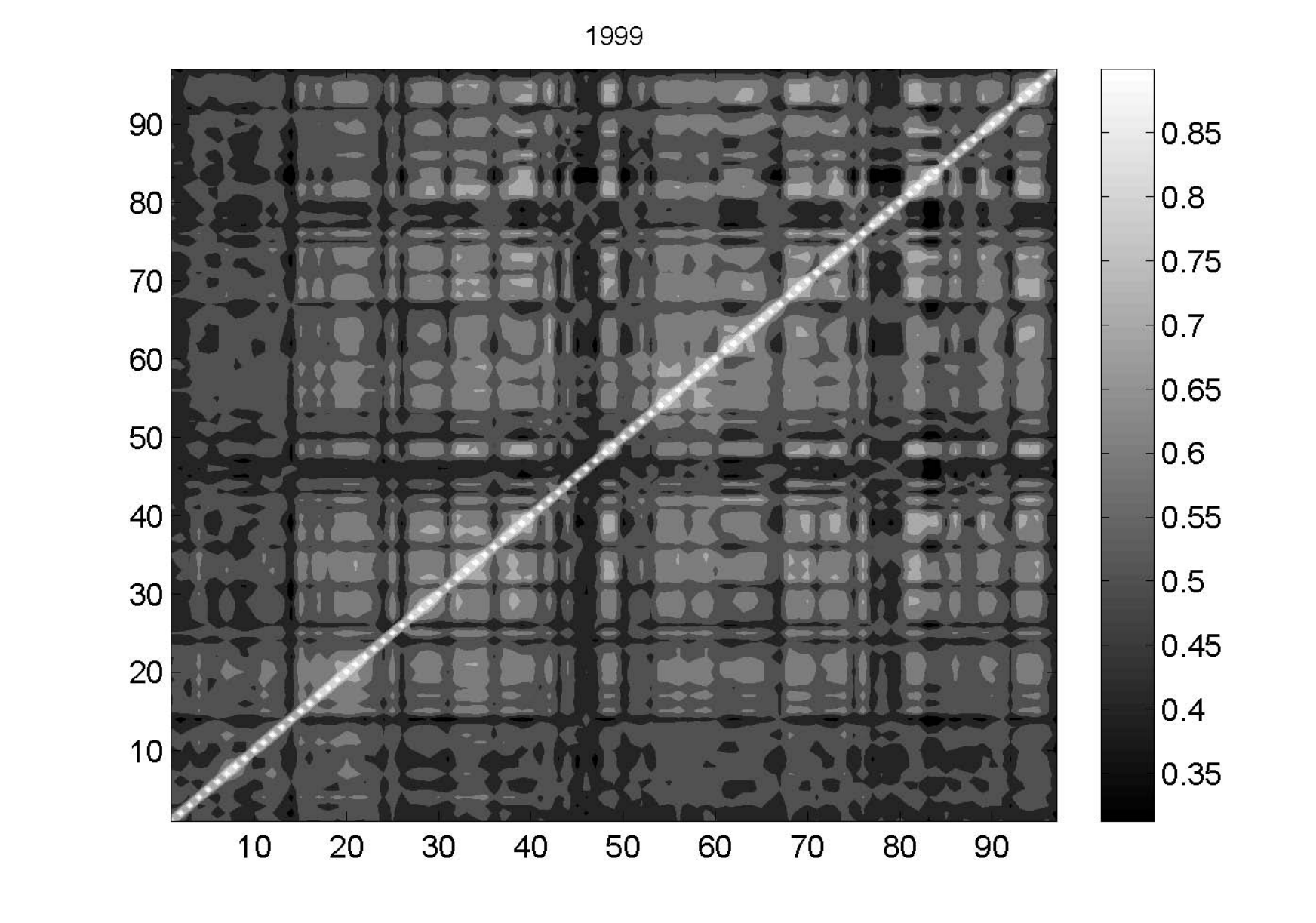}
\end{minipage}
\begin{minipage}{7cm}
\includegraphics[width=7cm]{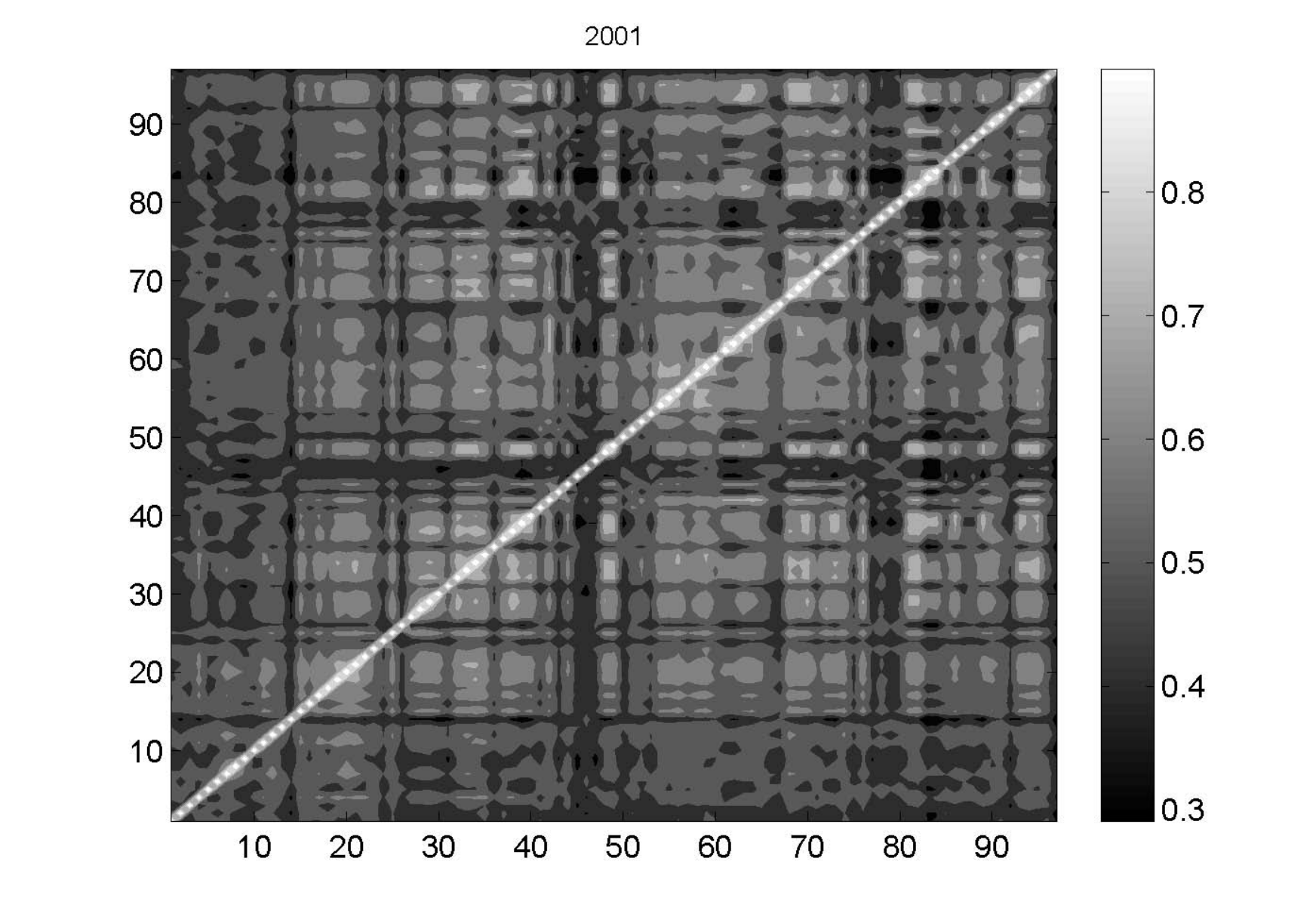}
\end{minipage}
\begin{minipage}{7cm}
\includegraphics[width=7cm]{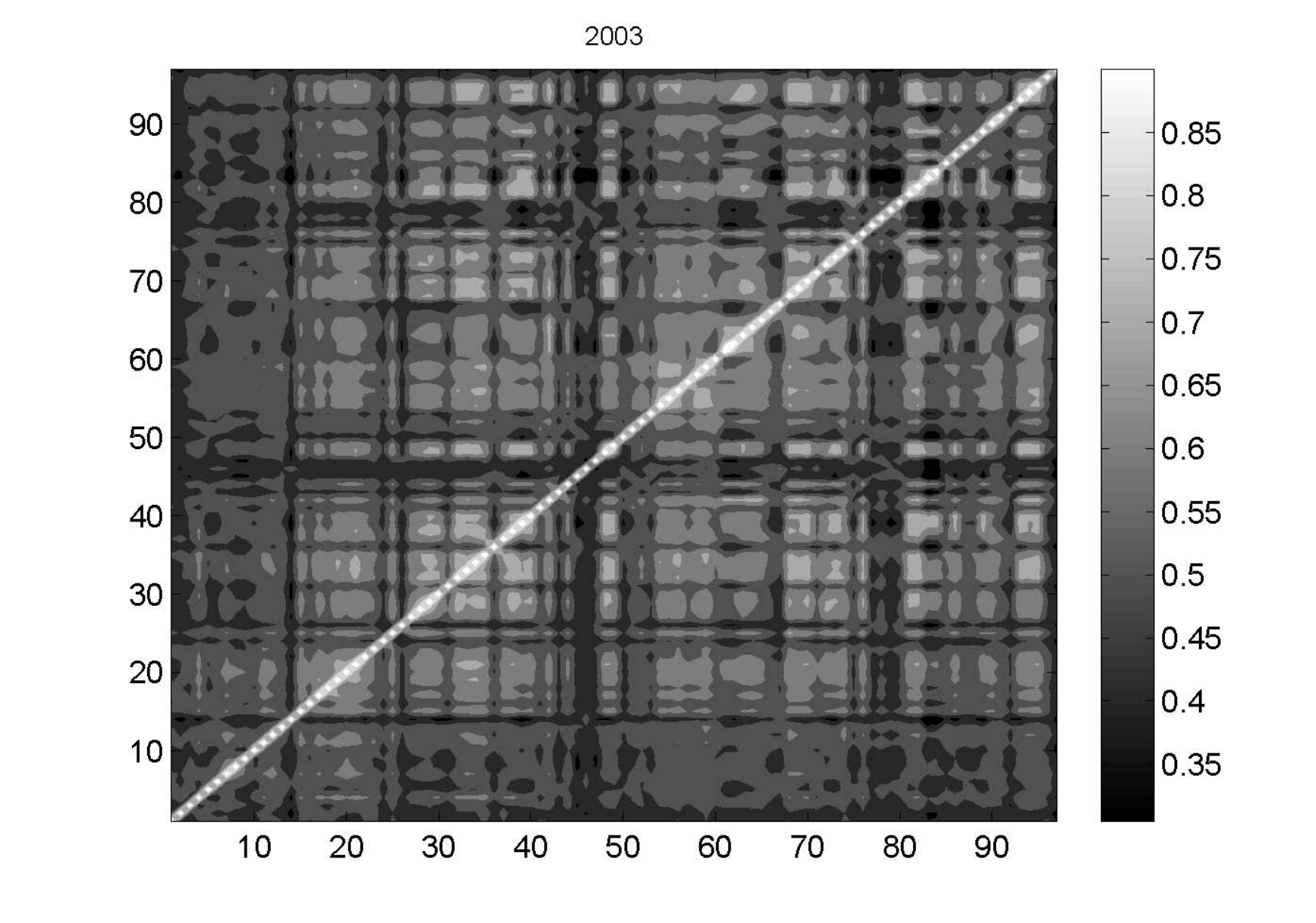}
\end{minipage}
\caption[]{\small Plots of unweighted inter-layer correlation
matrices $\Phi_u(t)$ for years $t=1993,1995,1997,1999,2001,2003$.
\label{fig_phiu}}
\end{center}
\end{sidewaysfigure}

Plots of the matrices $\Phi_w(t)$ and $\Phi_u(t)$ are shown for various years
in Figures \ref{fig_phiw} and \ref{fig_phiu} respectively. A first visual
inspection suggests that in both cases the observed correlation structure is
robust in time. However, as we show in section \ref{Subsec:sec_evol}, it is
possible to detect a small quantitative evolution of unweighted correlations,
and to interpret it as the manifestation of an underlying dynamics of trade
preferences determining ``revealed' correlations on top of ``intrinsic'' ones.
Before describing that effect, in the following section we discuss the result
of applying filtering procedures to inter-commodity correlation matrices.

\subsection{Hierarchies of layers}
The correlation matrices defined in Eqs. (\ref{eq_matrixphiw}) and
(\ref{eq_matrixphiu}) can be filtered exploiting a hierarchical procedure that
has been introduced in financial analysis \cite{Mantegna1999}. Starting from
the correlation coefficients $\phi^{c,c'}_w(t)$ or $\phi^{c,c'}_u(t)$ it is
possible to define a \emph{weighted/unweighted inter-layer distance} as
follows:
\begin{equation}\label{eq_distances}
d_{w/u}^{c,c'}(t)\equiv \sqrt{\frac{1-\phi_{w/u}^{c,c'}(t)}{2}}
\end{equation}
Notice that here we are introducing a normalized variant of the transformation
introduced in ref.\cite{Mantegna1999}. This has only an overall proportional
effect on all distances, and does not change their ranking or their metric
properties. We make this choice simply in order to have a maximum distance
value $d_{w/u}^{c,c'}=1$ when $c$ and $c'$ are perfectly anticorrelated
($\phi_{w/u}^{c,c'}=-1$), besides a minimum distance value $d_{w/u}^{c,c'}=0$
when $c$ and $c'$ are perfectly correlated ($\phi_{w/u}^{c,c'}=1$). One should
keep in mind that in case of no correlation ($\phi_{w/u}^{c,c'}=0$) the
above-defined distance equals $d_{w/u}^{c,c'}=1/\sqrt{2}\approx 0.707$.

Once a distance matrix is given, one can filter it to obtain a dendrogram
representing a taxonomy (hierarchical classification) of all layers. In such a
representation, the $C$ layers are the leaves of the taxonomic tree. Closer
(strongly correlated) layers meet at a branching point closer to the leaf
level, while more distant (weakly correlated) layers meet at a more distant
branching point. All layers eventually merge at a single root level. If the
tree is cut at some level, it splits in disconnected branches of similar (with
respect to the cut level chosen) layers. The hierarchical nature of the
classification is manifest in the nestedness of the dendrogram. A detailed
description of possible procedures to obtain the taxonomic tree can be found in
Ref. \cite{Mantegna1999}.

\begin{figure}  
\includegraphics[width=.7\textwidth]{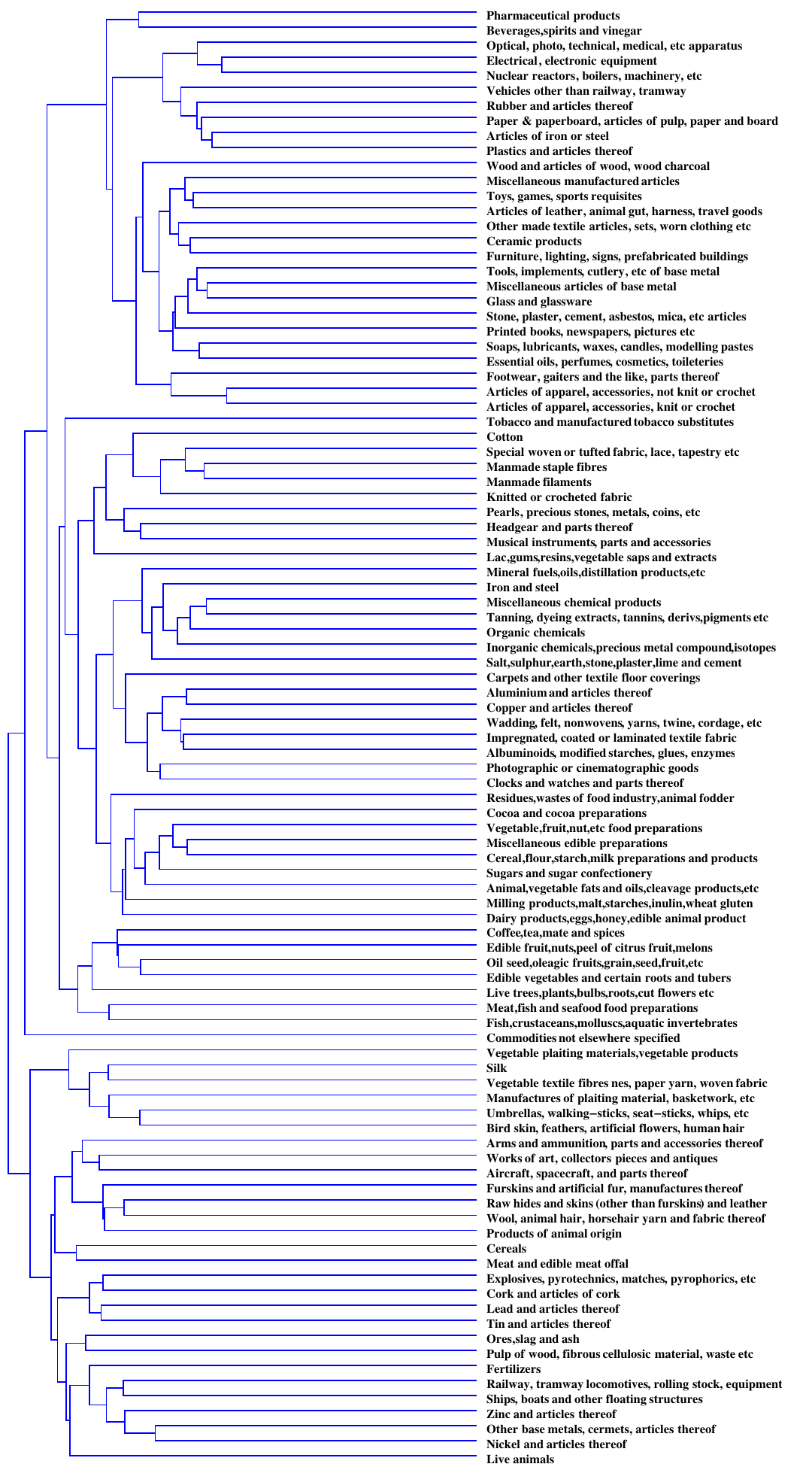}
\caption[]{\small Dendrogram of commodities obtained applying the Complete Linkage Clustering Algorithm to the unweighted inter-layer distances $d^{c,c'}_{u}(t)$ measured in year $t=2003$.
\label{fig_dendrogram1}}
\end{figure}

\begin{figure}  
\includegraphics[width=.7\textwidth]{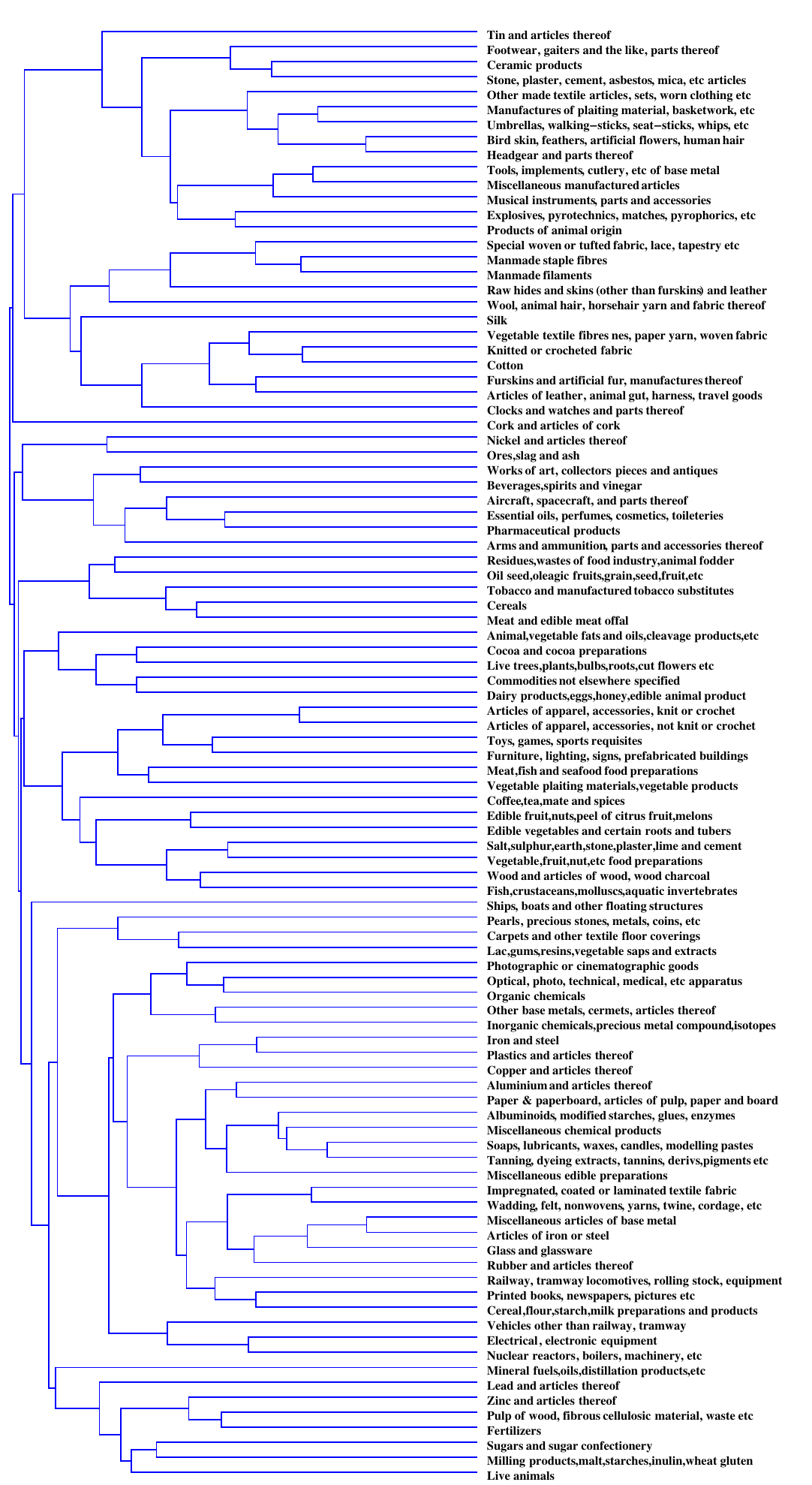}
\caption[]{\small Dendrogram of commodities obtained applying the Complete Linkage Clustering Algorithm to the weighted inter-layer distances $d^{c,c'}_{w}(t)$ measured in year $t=2003$.
\label{fig_dendrogram2}}
\end{figure}

In Figure \ref{fig_dendrogram1} we show the dendrogram of commodities obtained
applying the Complete Linkage Clustering Algorithm to the unweighted
inter-layer distances $d^{c,c'}_{u}(t)$ measured in year $t=2003$.
Similarly, in Figure \ref{fig_dendrogram2} we show we dendrogram obtained
applying the same algorithm to the weighted inter-layer distances
$d^{c,c'}_{w}(t)$ measured in the same year. In both dendrograms one can
observe that while in some cases similar commodities (such as the textiles and
leather sectors) are grouped together, in other cases a-priori unrelated goods
are found to belong to the same clusters. This confirms that, on top of an
intrinsic structure of inter-commodity correlations, ``revealed'' effects are
taking place. While it is not possible to disentangle these two contributions
on the basis of observed trade interactions alone, in the next section we
describe how we expect the two types of correlation to undergo different,
empirically observable, dynamical patterns.

\subsection{Evolution of inter-layer correlations and distances\label{Subsec:sec_evol}}
The previous results highlight that inter-commodity correlations are
a combination of ``revealed'' contributions, arising as
commodity-independent results of preferences in trade partnerships
between countries, and intrinsic contributions, due to inherent
commodity similarities. We now describe a way to assess whether
``revealed'' correlations develop in time on top of intrinsic
correlations. While the classification of trade commodities is
static (i.e. commodities do not become more or less similar as time
proceeds), the correlations among them may vary in time. This
implies that while intrinsic correlations are expected to remain
essentially stable in time as they merely reflect the internal
similarities already present in the commodity structure, revealed
correlation could in principle evolve in response of some dynamics
of trade preferences. Therefore we expect the time evolution of
inter-layer correlations and distances to reflect underlying
changes in trade preferences. Moreover, we expect trade preferences
to affect unweighted correlations more strongly than weighted
correlations, as they will primarily determine the presence or
absence of multiple types of traded commodities, while volumes will
be also affected by the specific sizes of production and demand.

We can study this effect in an aggregated fashion by defining the \emph{average
weighted/unweighted inter-layer correlation}
\begin{equation}
\bar{\phi}_{w/u}(t)\equiv \frac{\sum_{c\ne c'} \phi_{w/u}^{c,c'}(t)}{C(C-1)}=\frac{2 \sum_{c< c'} \phi_{w/u}^{c,c'}(t)}{C(C-1)}
\end{equation}
or, conversely, the \emph{average weighted/unweighted inter-layer distance}
\begin{equation}
\bar{d}_{w/u}(t)\equiv \frac{\sum_{c\ne c'} d_{w/u}^{c,c'}(t)}{C(C-1)}=\frac{2\sum_{c< c'} d_{w/u}^{c,c'}(t)}{C(C-1)}
\end{equation}
and following their evolution in time. Of course correlation and distance
measures are linked by \eqref{eq_distances}. Therefore, strictly speaking, the
only value added in studying them together is because they offer two
complementary interpretations of the same phenomenon.

\begin{figure}  
\includegraphics[width=.45\textwidth]{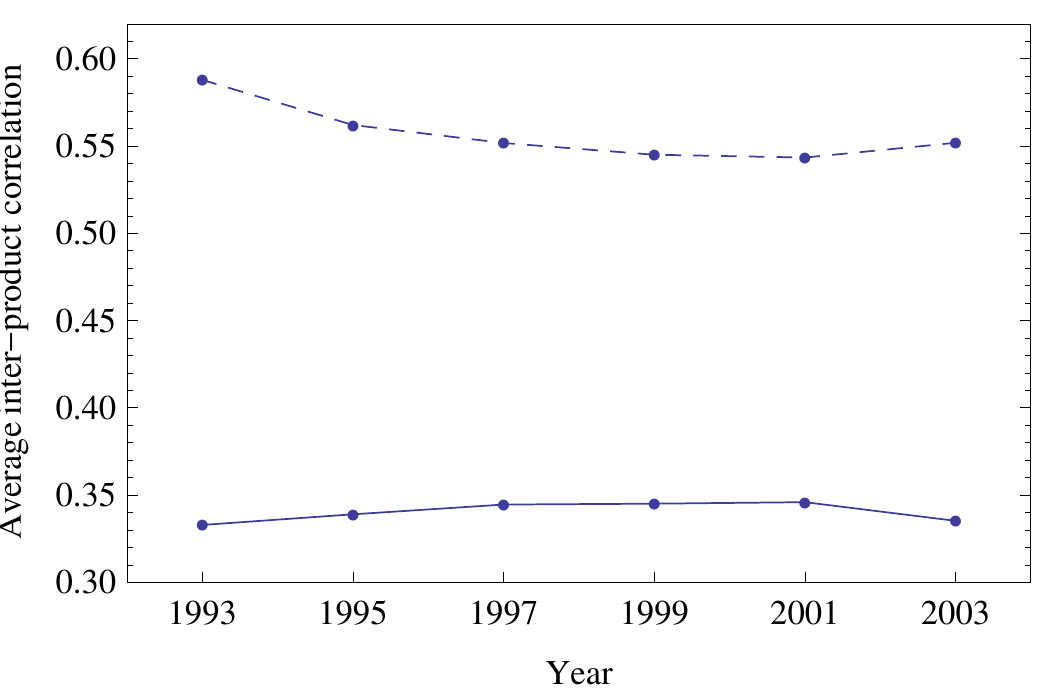}\hskip .1\textwidth\includegraphics[width=.45\textwidth]{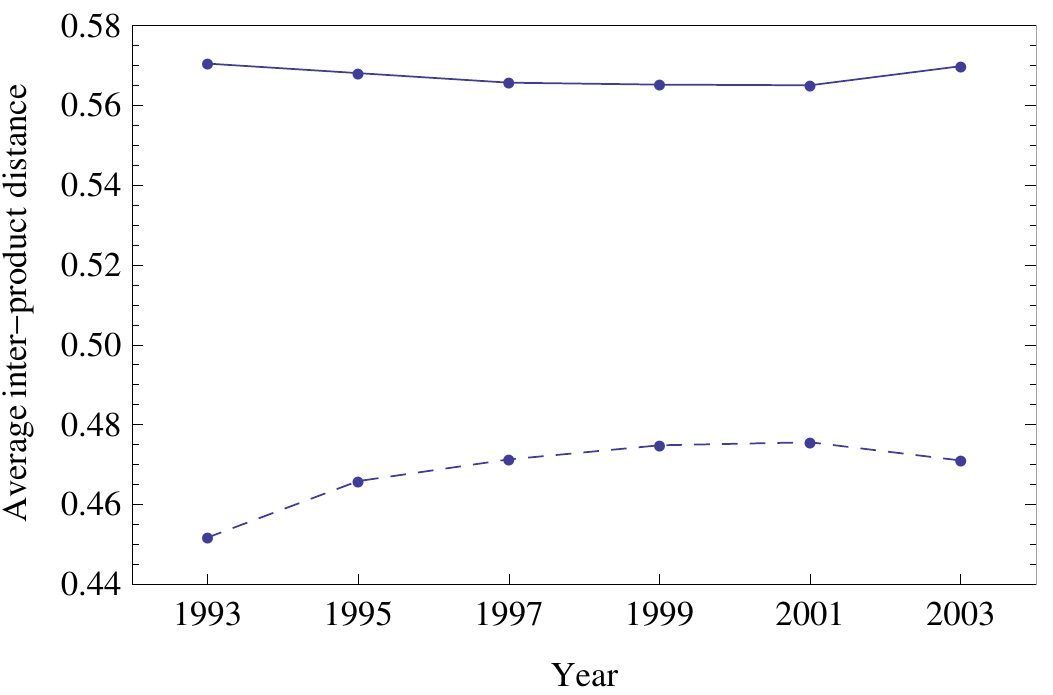}
\caption[]{\small Left: evolution of average weighted inter-layer correlation $\bar{\phi}_{w}(t)$ (solid) and average unweighted inter-layer correlation $\bar{\phi}_{u}(t)$ (dashed) from year $t=1993$ to year $t=2003$. Right: evolution of average weighted inter-layer distance $\bar{d}_{w}(t)$ (solid) and average unweighted inter-layer distance $\bar{d}_{u}(t)$ (dashed) from year $t=1993$ to year $t=2003$.
\label{fig_averagephi}}
\end{figure}

The results are shown in Figure \ref{fig_averagephi}. Note that the averages
are performed over all $C(C-1)/2$ commodity pairs. If all commodities were
uncorrelated one would have $\bar{\phi}_{w/u}=0$ and
$\bar{d}_{w/u}=1/\sqrt{2}$. The trends indicate that indeed a dynamics of
``revealed'' correlations is present. From year 1993 to year 2001, the average
unweighted inter-layer correlation $\bar{\phi}_{u}(t)$ has been decreasing
steadily over time, and correspondingly the average unweighted inter-layer
distance $\bar{d}_{u}(t)$ has been increasing. This means that, on average, the
roles played by different commodities in the international trade system have
become more and more dissimilar. The corresponding weighted quantities display
much smaller variations. We interpret these results as the enhancement of trade
specialization during the corresponding period, with pairs of countries
developing more and more commodity-intensive trade relationships characterized
by a decreasing variety of goods. As expected, this effect is more pronounced
for unweighted measures than for weighted measures, as the latter also
aggregate economy-specific size effects. However, from year 2001 to year 2003
an inversion in the trend is observed. Whether this is due to an actual
inversion of trade preferences is an important open point that requires further
clarification.

\section{Concluding remarks}\label{Sec:Conclusions}
In this paper we have begun to study the statistical properties of the
multi-network of international trade, and their evolution over time. We have
employed data on commodity-specific trade flows to build a sequence of graphs
where any two nodes (countries) are connected by many weighted directed edges,
each one representing the flow of export from the origin to the target country
for a given specific commodity class.

We have characterized the topological properties of all commodity-specific
networks and compared them to those of the aggregate-trade network.
Furthermore, we have studied both within- and across-network correlation
patterns between topological statistics, and tracked the time evolution of the largest connected components in the commodity-specific
networks. Finally, we have proposed a general approach to study multi-networks using detailed edge-level correlations among layers. This method allows to resolve the hierarchical organisation of inter-layer dependencies.
When applied to the trade network, it allows to define correlation-based inter-layer distances that are helpful in taxonomizing commodities not only
with respect to the inherent similarity between commodities, but also with
respect to the actual revealed trade patterns.

The preliminary nature of the present work opens the way to many possible
extensions. For instance, one might consider to employ filtering techniques
such as those use in Ref. \cite{serrc07} to extract in a multi-network
perspective a backbone of most-relevant trade-relationships between countries
that take into account, beside their geographical position and relative size,
also a third dimension defined by the type of commodities mostly traded.
Similarly, community detection techniques like the ones used in Ref.
\citep{Tzekina2008} may be extended to multi-network setups in order to single
out tightly-interconnected groups of countries, and possibly compare them to
the implications of international-trade models. Finally, the robustness of
statistical properties of the ITMNs might be checked against alternative
weighting schemes that, for example, control for country size and geographical
distance, much in the spirit of gravity models in international trade
literature \citep{Overman2003,Fratianni2009}.

\acknowledgements
D.G. acknowledges financial support from the European Commission 6th FP (Contract CIT3-CT-2005-513396), Project: DIME - Dynamics of Institutions and Markets in Europe.


\newpage

{\squeezetable \begin{longtable}{p{1cm}p{15cm}}
\caption[Harmonized System 1996 Classification of Commodities]{Harmonized System 1996 Classification of Commodities} \label{Tab:HS} \\
\hline \hline Code & Description \\ \hline
\endfirsthead

\hline \hline \multicolumn{2}{c}
{{\bfseries \tablename\ \thetable{} -- continued from previous page}} \\
Code & Description \\ \hline
\endhead

\hline \multicolumn{2}{r}{{Continued on next page}} \\
\endfoot

\hline \hline
\endlastfoot

\hline
        01 & Live animals \\

        02 & Meat and edible meat offal \\

        03 & Fish, crustaceans \& aquatic invertebrates \\

        04 & Dairy produce; birds eggs; honey and other edible animal products \\

        05 & Other products of animal origin \\

        06 & Live trees, plants; bulbs, roots; cut flowers \& ornamental foliage te \& spices \\

        07 & Edible vegetables \& certain roots \& Tubers \\

        08 & Edible fruit \& nuts; citrus fruit or melon peel  \\

        09 & Coffee, tea, mate \& spices  \\

        10 &   Cereals  \\

        11 & Milling products; malt; starch; inulin; wheat gluten \\

        12 & Oil seeds \& oleaginous fruits; miscellaneous grains, seeds \& fruit; industrial or medicinal plants; straw \& fodder  \\

        13 & Lac; gums, resins \& other vegetable sap \& extracts \\

        14 & Vegetable plaiting materials \& other vegetable products \\

        15 & Animal,vegetable fats and oils, cleavage products, etc \\

        16 & Edible preparations of meat, fish, crustaceans, molluscs or other aquatic invertebrates \\

        17 & Sugars and sugar confectionary  \\

        18 & Cocoa and cocoa preparations  \\

        19 & Preparations of cereals, flour, starch or milk; bakers wares  \\

        20 & Preparations of vegetables, fruit, nuts or other plant parts  \\

        21 & Miscellaneous edible preparations  \\

        22 & Beverages, spirits and vinegar  \\

        23 & Food industry residues \& waste; prepared animal feed  \\

        24 & Tobacco and manufactured tobacco substitutes  \\

        25 & Salt; sulfur; earth \& stone; lime \& cement plaster \\

        26 & Ores, slag and ash  \\

        27 & Mineral fuels, mineral oils \& products of their distillation; bitumin substances; mineral wax  \\

        28 & Inorganic chemicals; organic or inorganic compounds of precious metals, of rare-earth metals, of radioactive elements or of isotopes  \\

        29 & Organic chemicals  \\

        30 & Pharmaceutical products \\

        31 & Fertilizers  \\

        32 & Tanning or dyeing extracts; tannins \& derivatives; dyes, pigments \& coloring matter; paint \& varnish; putty \& other mastics; inks  \\

        33 & Essential oils and resinoids; perfumery, cosmetic or toilet preparations \\

        34 & Soap; waxes; polish; candles; modelling pastes; dental preparations with basis of plaster  \\

        35 & Albuminoidal substances; modified starch; glues; enzymes  \\

        36 & Explosives; pyrotechnic products; matches; pyrophoric alloys; certain combustible preparations  \\

        37 & Photographic or cinematographic goods  \\

        38 & Miscellaneous chemical products  \\

        39 & Plastics and articles thereof.  \\

        40 & Rubber and articles thereof. \\

        41 & Raw hides and skins (other than furskins) and leather  \\

        42 & Leather articles; saddlery and harness; travel goods, handbags \& similar; articles of animal gut [not silkworm gut]  \\

        43 & Furskins and artificial fur; manufactures thereof  \\

        44 & Wood and articles of wood; wood charcoal  \\

        45 & Cork and articles of cork  \\

        46 & Manufactures of straw, esparto or other plaiting materials; basketware \& wickerwork  \\

        47 & Pulp of wood or of other fibrous cellulosic material; waste \& scrap of paper \& paperboard  \\

        48 & Paper \& paperboard \& articles thereof; paper pulp articles ts and plans  \\

        49 & Printed books, newspapers, pictures and other products of printing industry; manuscripts, typescrip \\

        50 & Silk, including yarns and woven fabric thereof  \\

        51 & Wool \& animal hair, including yarn \& woven fabric  \\

        52 & Cotton, including yarn and woven fabric thereof  \\

        53 & Other vegetable textile fibers; paper yarn and woven fabrics of paper yarn  \\

        54 & Manmade filaments, including yarns \& woven fabrics  \\

        55 & Manmade staple fibres, including yarns \& woven fabrics  \\

        56 & Wadding, felt and nonwovens; special yarns; twine, cordage, ropes and cables and articles thereof  \\

        57 & Carpets and other textile floor coverings \\

        58 & Special woven fabrics; tufted textile fabrics; lace; tapestries; trimmings; embroidery \\

        59 & Impregnated, coated, covered or laminated textile fabrics; textile articles for industrial use  \\

        60 & Knitted or crocheted fabrics  \\

        61 & Apparel articles and accessories, knitted or crocheted  \\

        62 & Apparel articles and accessories, not knitted or crocheted  \\

        63 & Other textile articles; needlecraft sets; worn clothing and worn textile articles; rags \\

        64 & Footwear, gaiters and the like and parts thereof  \\

        65 & Headgear and parts thereof  \\

        66 & Umbrellas, walking-sticks, seat-sticks, riding-crops, whips, and parts thereof  \\

        67 & Prepared feathers, down and articles thereof; artificial flowers; articles of human hair \\

        68 & Articles of stone, plaster, cement, asbestos, mica or similar materials  \\

        69 & Ceramic products  \\

        70 & Glass and glassware \\

        71 & Pearls, precious stones, metals, coins, etc \\

        72 & Iron and steel  \\

        73 & Articles of iron or steel  \\

        74 & Copper and articles thereof  \\

        75 & Nickel and articles thereof  \\

        76 & Aluminum and articles thereof  \\

        78 & Lead and articles thereof  \\

        79 & Zinc and articles thereof  \\

        80 & Tin and articles thereof  \\

        81 & Other base metals; cermets; articles thereof \\

        82 & Tools, implements, cutlery, spoons \& forks of base metal \& parts thereof \\

        83 & Miscellaneous articles of base metal  \\

        84 & Nuclear reactors, boilers, machinery and mechanical appliances; parts thereof \\

        85 & Electric machinery, equipment and parts; sound equipment; television equipment  \\

        86 & Railway or tramway. Locomotives, rolling stock, track fixtures and parts thereof; mechanical \& electro-mechanical traffic signal equipment  \\

        87 & Vehicles, (not railway, tramway, rolling stock); parts and accessories \\

        88 & Aircraft, spacecraft, and parts thereof \\

        89 & Ships, boats and floating stuctures  \\

        90 & Optical, photographic, cinematographic, measuring, checking, precision, medical or surgical instruments/apparatus; parts \& accessories  \\

        91 & Clocks and watches and parts thereof  \\

        92 & Musical instruments; parts and accessories thereof  \\

        93 & Arms and ammunition, parts and accessories thereof \\

        94 & Furniture; bedding, mattresses, cushions etc; other lamps \& light fitting, illuminated signs and nameplates, prefabricated buildings  \\

        95 & Toys, games \& sports equipment; parts \& accessories \\

        96 & Miscellaneous manufactured articles  \\

        97 & Works of art, collectors pieces and antiques \\

        99 & Commodities not elsewhere specified \\
\hline
\end{longtable}}

\newpage \clearpage

\begin{sidewaystable}
\center \small
\begin{tabular}{p{1cm}p{6cm}p{2cm}p{2cm}p{2cm}}
\hline
\hline
   HS Code &  Commodity & Value (USD) & Value per Link (USD) & $\%$ of Aggregate Trade \\
\hline
        84 & Nuclear reactors, boilers, machinery and mechanical appliances; parts thereof  &   $5.67\times 10^{11}$ &   $6.17\times 10^{7}$ &    11.37\% \\

        85 & Electric machinery, equipment and parts; sound equipment; television equipment  &   $5.58\times 10^{11}$ &   $6.37\times 10^{7}$ &    11.18\% \\

        27 & Mineral fuels, mineral oils \& products of their distillation; bitumin substances; mineral wax  &   $4.45\times 10^{11}$ &   $9.91\times 10^{7}$ &     8.92\% \\

        87 & Vehicles, (not railway, tramway, rolling stock); parts and accessories  &   $3.09\times 10^{11}$ &   $4.76\times 10^{7}$ &     6.19\% \\

        90 & Optical, photographic, cinematographic, measuring, checking, precision, medical or surgical instruments/apparatus; parts \& accessories  &   $1.78\times 10^{11}$ &   $2.48\times 10^{7}$ &     3.58\% \\

        39 & Plastics and articles thereof.  &   $1.71\times 10^{11}$ &   $2.33\times 10^{7}$ &     3.44\% \\

        29 & Organic chemicals  &   $1.67\times 10^{11}$ &   $3.29\times 10^{7}$ &     3.35\% \\

        30 & Pharmaceutical products  &    $1.4\times 10^{11}$ &   $2.59\times 10^{7}$ &     2.81\% \\

        72 & Iron and steel  &   $1.35\times 10^{11}$ &   $2.77\times 10^{7}$ &     2.70\% \\

        71 & Pearls, precious stones, metals, coins, etc &   $1.01\times 10^{11}$ &   $2.41\times 10^{7}$ &     2.02\% \\

        10 &   Cereals  &   $3.63\times 10^{10}$ &   $1.28\times 10^{7}$ &     0.73\% \\

        52 & Cotton, including yarn and woven fabric thereof  &   $3.29\times 10^{10}$ &   $6.96\times 10^{6}$ &     0.66\% \\

         9 & Coffee, tea, mate \& spices  &   $1.28\times 10^{10}$ &   $2.56\times 10^{6}$ &     0.26\% \\

        93 & Arms and ammunition, parts and accessories thereof &   $4.31\times 10^{9}$ &   $2.46\times 10^{6}$ &     0.09\% \\

       ALL &  Aggregate &   $4.99\times 10^{12}$ &   $3.54\times 10^{8}$ &   100.00\% \\
       \hline
\hline
\end{tabular}
\caption{The 14 most relevant commodity classes in year 2003 in terms of
total-trade value (USD), trade value per link (USD), and share of world
aggregate trade.} \label{Tab:Top14}
\end{sidewaystable}

\begin{table}
\center \small
\begin{tabular}{llrrrrr}
\hline \hline
   HS Code &  Commodity &  $w_{ij}$ &    Density & $NS_{in}/ND_{in}$ & $NS_{out}/ND_{out}$ &     $WCC_{all}$ \\
\hline
         9 &     Coffee &      282\% &       27\% &      192\% &      177\% &      176\% \\

        10 &    Cereals &      497\% &       15\% &      540\% &      201\% &      218\% \\

        27 & Min. Fuels &      314\% &       24\% &      255\% &      282\% &      190\% \\

        29 & Org. Chem. &      277\% &       28\% &      218\% &      133\% &      176\% \\

        30 & Pharmaceutical &      260\% &       29\% &      248\% &      111\% &      151\% \\

        39 &   Plastics &      192\% &       40\% &      173\% &      107\% &      119\% \\

        52 &     Cotton &      298\% &       26\% &      227\% &      162\% &      220\% \\

        71 & Prec. Metals &      337\% &       23\% &      192\% &      206\% &      151\% \\

        72 &       Iron &      290\% &       26\% &      243\% &      145\% &      182\% \\

        84 & Nuclear Machin. &      153\% &       50\% &      140\% &      101\% &      109\% \\

        85 & Electric Machin. &      161\% &       48\% &      139\% &      102\% &      109\% \\

        87 &   Vehicles &      217\% &       35\% &      201\% &      106\% &      115\% \\

        90 & Optical Instr. &      196\% &       39\% &      153\% &      104\% &      112\% \\

        93 &       Arms &      804\% &       10\% &      576\% &      350\% &      375\% \\
\hline \hline
\end{tabular}
\caption{Density and node-average of topological properties of
commodity-specific networks vs. aggregate trade network for the 14 most
relevant commodity classes in year 2003. Percentages refer to the ratio of the
statistic value in the commodity-specific network to aggregate network. Values
larger (smaller) than 100\% mean that average of commodity-specific networks is
larger (smaller) than its counterpart in the aggregate network.}
\label{Tab:Averages}
\end{table}

\begin{table}
\center \small
\begin{tabular}{rrrrrr}
\hline
\hline
           &            &           \multicolumn{ 4}{c}{Clustering Pattern} \\

   HS Code &  Commodity &      Cycle &  Middleman &         In &        Out \\
\hline
        09 & Coffee and spices  &     2.77\% &    18.81\% &    34.92\% &    43.50\% \\

        10 &   Cereals  &     2.19\% &    14.86\% &    57.93\% &    25.02\% \\

        27 & Mineral fuels &     3.13\% &    20.66\% &    39.18\% &    37.03\% \\

        29 & Organic chemicals  &     8.94\% &    11.06\% &    49.47\% &    30.53\% \\

        30 & Pharmaceutical products  &     4.93\% &     6.13\% &    64.79\% &    24.15\% \\

        39 &   Plastics &     7.73\% &    10.52\% &    51.54\% &    30.21\% \\

        52 &     Cotton &     7.71\% &    12.94\% &    44.13\% &    35.22\% \\

        71 & Precious metals &    14.00\% &    15.84\% &    17.72\% &    52.44\% \\

        72 & Iron and steel  &     7.13\% &    15.40\% &    45.28\% &    32.20\% \\

        84 & Nuclear machinery &     7.77\% &     9.46\% &    51.88\% &    30.89\% \\

        85 & Electric machinery &     9.27\% &    10.33\% &    48.15\% &    32.26\% \\

        87 &   Vehicles &     5.49\% &     7.45\% &    57.48\% &    29.58\% \\

        90 & Optical instruments &     9.10\% &    10.63\% &    48.39\% &    31.88\% \\

        93 &       Arms &     6.69\% &    13.74\% &    54.68\% &    24.90\% \\

       All &  Aggregate &    20.21\% &    20.69\% &    22.46\% &    36.64\% \\
\hline \hline
\end{tabular}
\caption{Relative frequency of the occurrence of clustering patterns in the
aggregate and commodity-specific networks.} \label{Tab:ClustPerc}
\end{table}

\begin{table}
\begin{tabular}{lrrrrrrr}
\hline
\hline
      Area &          N &       1993 &       1995 &       1997 &       1999 &       2001 &       2003 \\
\hline
   Core EU &          8 &       63\% &      100\% &      100\% &      100\% &      100\% &      100\% \\

Periphery EU &         10 &       90\% &      100\% &      100\% &      100\% &      100\% &      100\% \\

Eastern Europe &         15 &       20\% &       53\% &       93\% &      100\% &       93\% &       93\% \\

North and Central America &         22 &       59\% &       73\% &       91\% &       95\% &       91\% &       82\% \\

South America &         12 &       58\% &       92\% &       83\% &      100\% &      100\% &       83\% \\

South and East Asia &         20 &       65\% &       55\% &       65\% &       70\% &       75\% &       80\% \\

Central Asia &          8 &       13\% &       25\% &       50\% &       50\% &       38\% &       63\% \\

North Africa and Middle East &         18 &       39\% &       56\% &       56\% &       61\% &       78\% &       78\% \\

Sub-Saharan Africa &         40 &       18\% &       58\% &       65\% &       70\% &       70\% &       53\% \\

   Oceania &          9 &       33\% &       33\% &       33\% &       33\% &       44\% &       56\% \\

     World &        162 &       41\% &       63\% &       72\% &       77\% &       79\% &       74\% \\
\hline \hline
\end{tabular}

\center \small \caption{Size of the largest connected component as a percentage
of total network size across geographical macro-areas and time in the aggregate
(all-commodity) trade network. Here two nodes are said to be connected if they
are linked by a bilateral edge (both import and export relationship). }
\label{Tab:ConnCompAggr}
\end{table}

\begin{table}
\begin{tabular}{rlrccc}
\hline
\hline
   HS Code &  Commodity &        All & Largest 10\% & Largest 5\% & Largest 1\% \\
\hline
      09   & Coffee and spices  &        119 &         46 &         23 &          4 \\

      10   &   Cereals  &        107 &         25 &         15 &          3 \\

        27 & Mineral fuels &        117 &         45 &         28 &          9 \\

        29 & Organic chemicals  &        117 &         41 &         29 &         11 \\

        30 & Pharmaceutical products  &        117 &         40 &         23 &         10 \\

        39 &   Plastics &        120 &         57 &         40 &         19 \\

        52 &     Cotton &        116 &         45 &         29 &         12 \\

        71 & Precious metals &        114 &         42 &         27 &         11 \\

        72 & Iron and steel  &        119 &         45 &         33 &         14 \\

        84 & Nuclear machinery &        120 &         45 &         39 &         21 \\

        85 & Electric machinery &        120 &         48 &         39 &         19 \\

        87 &   Vehicles &        120 &         46 &         34 &         14 \\

        90 & Optical instruments &        120 &         48 &         33 &         14 \\

        93 &       Arms &         80 &         23 &         17 &          5 \\

       All &  Aggregate &        162 &         81 &         58 &         28 \\
\hline \hline
\end{tabular}
\center \small \caption{Size of the largest connected component in aggregate
and commodity-specific networks in year 2003. All: A binary link is in place if
the associated link weight is larger than zero; Largest x\%: A binary link is
in place if the associated link weight belongs to the set of x\% largest
link-weights. Here we assume that two nodes are connected if either an inward
or an outward link is in place.} \label{Tab:ConnCompDisaggr}
\end{table}

{\squeezetable
\begin{table} \centering
\begin{tabular}{lllp{6cm}} \hline \hline
   HS Code &  Commodity & Size of LCC & Countries in the LCC \\
\hline
      09   & Coffee and spices  &          4 & Canada; Germany; Italy; USA \\

      10   &   Cereals  &          3 & Canada; Germany; USA \\

        27 & Mineral fuels &          9 & Canada; China; Germany; Indonesia; Korea; Malaysia; Singapore; UK; USA \\

        29 & Organic chemicals  &         11 & Canada; China; France; Germany; Italy; Japan; Korea; Netherlands; Switzerland; UK; USA \\

        30 & Pharmaceutical products  &         10 & Canada; France; Germany; Italy; Japan; Netherlands; Spain;  Switzerland; UK; USA \\

        39 &   Plastics &         19 & Austria; Canada; China; France; Germany; Hong Kong; Italy; Japan; Korea; Malaysia; Mexico; Netherlands; Poland; Singapore; Spain; Switzerland; Thailand; UK; USA \\

        52 &     Cotton &         12 & China; France; Germany; Hong Kong; Italy; Japan; Korea; Mexico;  Pakistan; Spain; Turkey; USA \\

        71 & Precious metals &         11 & Australia; Belgium-Luxembourg; Canada; Hong Kong; India; Israel;  Italy; Korea; Switzerland; UK; USA \\

        72 & Iron and steel  &         14 & Austria; Canada; China; France; Germany; Italy; Japan; Korea Mexico; Netherlands; Russia; Spain; UK; USA  \\

        84 & Nuclear machinery &         21 & Austria; Brazil; Canada; China; France; Germany; Ireland; Italy; Japan; Korea; Malaysia; Mexico; Netherlands; Philippines; Poland; Singapore; Spain; Sweden; Thailand; UK; USA \\

        85 & Electric machinery &         19 & Austria; Canada; China; France; Germany; Hong Kong; Hungary; Italy; Japan; Korea; Malaysia; Mexico; Netherlands; Philippines; Singapore; Switzerland; Thailand; UK; USA \\

        87 &   Vehicles &         14 & Canada; China; France; Germany; Hungary; Italy; Japan; Mexico; Netherlands; Poland; Spain; Sweden; UK; USA \\

        90 & Optical instruments &         14 & Canada; China; France; Germany; Hong Kong; Ireland; Italy; Japan; Mexico; Netherlands; Singapore; Switzerland; UK; USA \\

        93 &       Arms &          5 & Canada; Italy; Japan; Spain; USA \\

       All &  Aggregate &         28 & Australia; Austria; Brazil; Canada; China; Denmark; France Germany; Hong Kong; Hungary; Ireland; Italy; Japan; Korea; Malaysia; Mexico; Netherlands; Philippines; Poland; Russia Singapore; Spain; Sweden; Switzerland; Thailand; Turkey UK; USA
 \\
\hline \hline \end{tabular} \center \small \caption{Size and composition of the
largest-connected component (LCC) in aggregate and commodity-specific networks
in year 2003. A binary link is in place if the associated link weight belongs
to the set of 1\% largest link-weights. Here we assume that two nodes are
connected if either an inward or an outward link is in place.}
\label{Tab:ConnCompComposition}
\end{table}}

\begin{sidewaystable}
\centering \small
\begin{tabular}{rccccccccc}
\hline
\hline
           &      \multicolumn{ 3}{c}{$NS_{in}$} &     \multicolumn{ 3}{c}{$NS_{out}$} &     \multicolumn{ 3}{c}{$NS_{tot}$} \\

 Commodity &        1st &        2nd &        3rd &        1st &        2nd &        3rd &        1st &        2nd &        3rd \\
\hline
Coffee \& spices  &        USA &    Germany &      Japan &     Brazil &   Colombia &  Indonesia &        USA &    Germany &     Brazil \\

  Cereals  &      Japan &     Mexico &      Korea &        USA &     France &  Argentina &        USA &      Japan &     France \\

Mineral fuels &        USA &      Japan &      China &     Russia & Saudi Arabia &     Norway &        USA &     Russia &      China \\

Organic chemicals  &        USA &      China &    Germany &        USA &    Ireland &    Germany &        USA &    Germany &     France \\

Pharmaceutical products  &        USA &    Germany &         UK &        USA &    Germany &     France &        USA &    Germany &         UK \\

  Plastics &      China &        USA &    Germany &    Germany &        USA &      Japan &    Germany &        USA &      China \\

    Cotton &  Hong Kong &      China &        USA &      China &        USA &      Italy &      China &        USA &  Hong Kong \\

Precious metals &        USA &  Hong Kong &         UK & Switzerland &      India &        USA &        USA &      India & Switzerland \\

Iron \& steel  &      China &        USA &      Italy &      Japan &    Germany &     Russia &    Germany &      China &      Japan \\

Nuclear machinery &        USA &         UK &    Germany &    Germany &        USA &      China &        USA &    Germany &      China \\

Electric machinery &    Germany &        USA &         UK &        USA &      China &    Germany &        USA &    Germany &      China \\

  Vehicles &    Germany &        USA &         UK &    Germany &      Japan &     France &    Germany &      Japan &         UK \\

Optical instruments &        USA &    Germany &         UK &        USA &    Germany &      Japan &        USA &    Germany &      Japan \\

      Arms &        USA &         UK &      Korea &        USA &    Germany &      Italy &        USA &         UK &    Germany \\

 Aggregate &        USA &    Germany &         UK &        USA &    Germany &      China &        USA &    Germany &      China \\
\hline \hline
\end{tabular}
\caption{Country rankings in 2003. Top 3 position according to node strength
statistics.} \label{Tab:RankNS}
\end{sidewaystable}

{\squeezetable
\begin{sidewaystable} \centering
\begin{tabular}{rrrrrrr}
\hline
\hline
           &    \multicolumn{ 3}{c}{$NS_{in-in}$} &   \multicolumn{ 3}{c}{$NS_{out-in}$} \\

 Commodity &        1st &        2nd &        3rd &        1st &        2nd &        3rd \\
\hline
Coffee \& spices  &   Cambodia &   Dominica &     Guyana & St. Kitts \& Nevis & Eq. Guinea &    Vanuatu \\

  Cereals  & Sao Tome & Papua New Guinea &      Samoa &   Mongolia &      Nepal &    Morocco \\

Mineral fuels & C. African Rep &      Samoa &    Grenada & Guinea Bissau & Mauritania & DemRepCongo \\

Organic chemicals  & C. African Rep &     Gambia & St. Vincent &    Vanuatu &     Guyana & St. Vincent \\

Pharmaceutical products  & C. African Rep &      Samoa &   Maldives &    Bahamas &      Nepal & Kyrgyzstan \\

  Plastics & C. African Rep &      Samoa &   Maldives &    Comoros & Eq. Guinea &   Mongolia \\

    Cotton &  St. Lucia &     Belize &   Mongolia &   Mongolia &       Laos &     Malawi \\

Precious metals &      Gabon & St. Kitts \& Nevis &     Gambia & Cape Verde &   St.Lucia & Eq. Guinea \\

Iron \& steel  &      Samoa &      Nepal &    Grenada &   Mongolia & Papua New Guinea & Sao Tome \\

Nuclear machinery & Sao Tome &      Samoa & Brunei  Darussalam & Eq. Guinea &     Rwanda & C. African Rep \\

Electric machinery & C. African Rep &      Samoa & Sao Tome &   Kiribati &   Djibouti &   Mongolia \\

  Vehicles & St. Kitts \& Nevis & Sao Tome &   Dominica &   Suriname & Sao Tome & Cape Verde \\

Optical instruments &      Samoa & C. African Rep &     Gambia & St. Vincent &      Haiti & Cape Verde \\

      Arms & St. Kitts \& Nevis & Papua New Guinea &   Dominica &    Ecuador &      Haiti &    Albania \\

 Aggregate & Sao Tome &      Samoa &   Maldives &      Tonga &   St.Lucia &   Kiribati \\
\hline \hline
           &            &            &            &            &            &            \\

           &            &            &            &            &            &            \\
\hline \hline
           &  \multicolumn{ 3}{c}{$NS_{out-out}$} &   \multicolumn{ 3}{c}{$NS_{in-out}$} \\

 Commodity &        1st &        2nd &        3rd &        1st &        2nd &        3rd \\
\hline
Coffee \& spices  &     Bhutan & St. Kitts \& Nevis &       Chad &     Guyana & ElSalvador &    Ecuador \\

  Cereals  &    Armenia &     Bhutan &    Jamaica & C. African Rep &      Samoa &     Guyana \\

Mineral fuels &   Mongolia & Tajikistan & Kyrgyzstan &  Hong Kong &      Gabon &     Rwanda \\

Organic chemicals  & St. Vincent & Tajikistan & Eq. Guinea &     Gambia & C. African Rep &   Cambodia \\

Pharmaceutical products  &    Bahamas &   Suriname &      Nepal & C. African Rep &      Samoa &     Gambia \\

  Plastics & St. Kitts \& Nevis &    Comoros &    Grenada & C. African Rep &      Samoa &   Maldives \\

    Cotton &   Mongolia &    Bahamas &     Gambia &   St.Lucia &   Dominica &     Belize \\

Precious metals &   Kiribati &     Uganda & Cape Verde &     Guyana &      Samoa &     Malawi \\

Iron \& steel  & Sao Tome & Madagascar & Sierra Leone &   Mongolia & Brunei Darussalam &      Nepal \\

Nuclear machinery & Eq. Guinea & Cape Verde & St. Kitts \& Nevis & Sao Tome &      Samoa & St. Kitts \& Nevis \\

Electric machinery &   Kiribati & Tajikistan &   Mongolia & Sao Tome &      Samoa &     Belize \\

  Vehicles &   Suriname & SolomonIsds & Sao Tome & Sao Tome & St. Kitts \& Nevis &   Dominica \\

Optical instruments & St. Vincent &      Haiti & Cape Verde & C. African Rep &      Samoa &     Gambia \\

      Arms &    Ecuador &      Haiti &    Albania & St. Kitts \& Nevis & Papua New Guinea &   Dominica \\

 Aggregate &      Tonga &   St.Lucia &     Bhutan & Sao Tome &      Samoa &   Maldives \\
\hline \hline
\end{tabular}
 \caption{Country rankings in 2003. Top 3 position according
to node ANNS statistics.} \label{Tab:RankANNS}
\end{sidewaystable}}

{\squeezetable
\begin{table} \centering
\begin{tabular}{rcccccc}
\hline
\hline
           &    \multicolumn{ 3}{c}{$WCC_{all}$} &       \multicolumn{ 3}{c}{$WCENTR$} \\

 Commodity &        1st &        2nd &        3rd &        1st &        2nd &        3rd \\
\hline
Coffee \& spices  &   Colombia &     Brazil &    Vietnam &     Brazil &   Colombia &  Indonesia \\

  Cereals  &    Algeria & Papua New Guinea &    Tunisia &        USA &     Canada &   Thailand \\

Mineral fuels & Eq. Guinea &      Libya &     Angola &     Russia & Saudi Arabia &     Norway \\

Organic chemicals  & Eq. Guinea &        USA &      Japan &    Ireland &        USA &    Germany \\

Pharmaceutical products  &        USA &    Germany &     France &        USA &    Germany &     France \\

  Plastics &    Germany &        USA &      China &    Germany &        USA & Netherlands \\

    Cotton & Uzbekistan &      China &      Italy &      China &        USA &   Pakistan \\

Precious metals &     Israel & Uzbekistan &     Angola & Switzerland &      India &         UK \\

Iron \& steel  &    Germany &      Italy &      China &    Germany &     France &      Japan \\

Nuclear machinery &      China &        USA &    Germany &      China &      Japan &        USA \\

Electric machinery &      China &        USA &    Germany &        USA &      Japan &      China \\

  Vehicles &    Germany &      Japan &        USA &    Germany &      Japan &         UK \\

Optical instruments &        USA &      China &      Japan &        USA &      China &      Japan \\

      Arms & Saudi Arabia &     Norway &        USA &        USA &    Germany &      Italy \\

 Aggregate &        USA &    Germany &      China &        USA &      China &    Germany \\
\hline \hline
\end{tabular}
\caption{Country rankings in 2003. Top 3 position according to node overall
clustering and centrality statistics.} \label{Tab:RankWCC_CENTR}
\end{table}}

{\squeezetable \begin{sidewaystable} \centering
\begin{tabular}{rrccccccccccc}
\hline
\hline
           &            &                                                                                                \multicolumn{ 11}{c}{Correlation Coefficient} \\

           &            & $NS_{in}$ & $NS_{out}$ & $ANNS_{tot}$ & $ANNS_{in-in}$ & $ANNS_{in-out}$ & $ANNS_{out-in}$ & $ANNS_{out-out}$ & $WCC_{all}$ & $WCC_{in}$ & $WCC_{out}$ &  $WCENTR$ \\

   HS Code &  Commodity & $ND_{in}$ & $ND_{out}$ & $NS_{tot}$ & $NS_{in}$ & $NS_{in}$ & $NS_{out}$ & $NS_{out}$ & $NS_{tot}$ & $NS_{in}$ & $NS_{out}$ & $NS_{tot}$ \\
\hline
        09 & Coffee \& spices  &     0.5916 &     0.6311 &    -0.3511 &    -0.0922 &    -0.0527 &    -0.2666 &    -0.0777 &     0.6462 &     0.7485 &     0.7283 &     0.6247 \\

        10 &   Cereals  &     0.4663 &     0.6454 &    -0.1151 &     0.1704 &     0.0592 &    -0.0119 &    -0.0522 &     0.3130 &     0.7328 &     0.5663 &     0.7957 \\

        27 & Mineral fuels &     0.6615 &     0.4937 &    -0.1746 &    -0.0474 &     0.1631 &    -0.0208 &     0.0121 &     0.3629 &     0.8605 &     0.5195 &     0.7295 \\

        29 & Organic chemicals  &     0.5256 &     0.6242 &    -0.2428 &    -0.0918 &    -0.0808 &    -0.1721 &    -0.1583 &     0.7810 &     0.8484 &     0.7227 &     0.9116 \\

        30 & Pharmaceutical products  &     0.4642 &     0.5876 &    -0.2123 &    -0.0053 &    -0.0266 &    -0.1489 &    -0.1489 &     0.9148 &     0.7677 &     0.9681 &     0.9702 \\

        39 &   Plastics &     0.5828 &     0.5376 &    -0.3610 &    -0.0452 &    -0.0672 &    -0.2942 &    -0.2990 &     0.9148 &     0.7721 &     0.9600 &     0.9667 \\

        52 &     Cotton &     0.6226 &     0.6455 &    -0.3280 &    -0.0921 &    -0.1310 &    -0.1845 &    -0.1849 &     0.5967 &     0.7668 &     0.5322 &     0.8843 \\

        71 & Precious metals &     0.6263 &     0.6775 &    -0.3437 &    -0.1531 &    -0.1328 &    -0.2790 &    -0.3125 &     0.6860 &     0.7624 &     0.6691 &     0.9097 \\

        72 & Iron and steel  &     0.5478 &     0.7140 &    -0.3694 &    -0.0323 &    -0.0139 &    -0.2158 &    -0.2081 &     0.8386 &     0.8798 &     0.7900 &     0.8559 \\

        84 & Nuclear machinery &     0.6630 &     0.5618 &    -0.5377 &    -0.0676 &    -0.0948 &    -0.4667 &    -0.4511 &     0.9323 &     0.7680 &     0.9782 &     0.9567 \\

        85 & Electric machinery &     0.6431 &     0.5916 &    -0.5069 &    -0.1002 &    -0.1122 &    -0.4753 &    -0.4526 &     0.9327 &     0.7927 &     0.9752 &     0.9494 \\

        87 &   Vehicles &     0.5938 &     0.5165 &    -0.3498 &    -0.0150 &    -0.0635 &    -0.2659 &    -0.2440 &     0.9171 &     0.7435 &     0.9612 &     0.9746 \\

        90 & Optical instruments &     0.6134 &     0.4819 &    -0.3634 &    -0.1299 &    -0.1400 &    -0.3173 &    -0.2868 &     0.9105 &     0.7414 &     0.9588 &     0.9564 \\

        93 &       Arms &     0.5948 &     0.6956 &    -0.1215 &    -0.0422 &    -0.0476 &    -0.0553 &    -0.0659 &     0.5374 &     0.7078 &     0.4825 &     0.8358 \\

       All &  Aggregate &     0.4453 &     0.4620 &    -0.4017 &    -0.1437 &    -0.1412 &    -0.4348 &    -0.4377 &     0.9669 &     0.9494 &     0.9760 &     0.9779 \\
\hline \hline
\end{tabular}
\caption{Correlation coefficients between topological statistics within each
commodity-network in year 2003.} \label{Tab:CorrWithin}
\end{sidewaystable}}



\clearpage

\newpage \clearpage

\newpage \clearpage

\newpage \clearpage

\newpage \clearpage

\newpage \clearpage

\end{document}